\documentclass[12pt]{iopart}

\usepackage{graphicx,hyperref,iopams}
\usepackage[square,sort&compress,numbers]{natbib}
\usepackage[varg]{txfonts}

\newcommand{\D}{\mathcal{D}}
\newcommand{\cvector}[1]{\left(\begin{array}{c}#1\end{array}\right)}
\renewcommand{\matrix}[2]{\left(\begin{array}{#1}#2\end{array}\right)}
\renewcommand{\bi}[1]{\boldsymbol{#1}}

\begin{document}

\title
  [Microscopic field theory for cosmic structures]
  {A microscopic, non-equilibrium, statistical field theory for cosmic structure formation}
\author{Matthias Bartelmann, Felix Fabis, Daniel Berg, Elena Kozlikin, Robert Lilow, Celia Viermann}
\address{Heidelberg University, Zentrum f\"ur Astronomie, Institut f\"ur Theoretische Astrophysik, Philosophenweg 12, 69120 Heidelberg, Germany}
\ead{bartelmann@uni-heidelberg.de}

\begin{abstract}
Building upon the recent pioneering work by Mazenko and by Das \& Mazenko, we develop a microscopic, non-equilibrium, statistical field theory for initially correlated canonical ensembles of classical microscopic particles obeying Hamiltonian dynamics. Our primary target is cosmic structure formation, where initial Gaussian correlations in phase space are believed to be set by inflation. We give an exact expression for the generating functional of this theory and work out suitable approximations. We specify the initial correlations by a power spectrum and derive general expressions for the correlators of the density and the response field. We derive simple closed expressions for the lowest-order contributions to the non-linear cosmological power spectrum, valid for arbitrary wave numbers. We further calculate the bispectrum expected in this theory within these approximations and the power spectrum of cosmic density fluctuations to first order in the gravitational interaction, using a recent improvement of the Zel'dovich approximation. We show that, with a modification motivated by the adhesion approximation, the non-linear growth of the density power spectrum found in numerical simulations of cosmic structure evolution is reproduced well to redshift zero and for arbitrary wave numbers even within first-order perturbation theory. Our results present the first fully analytic calculation of the non-linear power spectrum of cosmic structures.
\end{abstract}

\pacs{04.40.-b, 05.20.-y, 98.65.Dx}

\noindent{\it Keywords\/}: nonequilibrium dynamics, self-gravitating systems, cosmic structure formation

\submitto{\NJP}

% \maketitle

\section{Introduction}
\label{sc:1}

\subsection{Motivation and overview}

In a sequence of pioneering papers, Mazenko and Das \& Mazenko \cite{2013JSP...152..159D, 2012JSP...149..643D, 2011PhRvE..83d1125M, 2010PhRvE..81f1102M} have recently shown how the non-equilibrium kinetic theory of classical particles can be mapped to the path-integral approach familiar from statistical quantum field theory, in the spirit of \cite{1973PhRvA...8..423M} and \cite{1977PhRvA..16..732F}. Besides the unifying formal analogy, this approach has a multitude of advantages for the systematic development of perturbation theory and the calculation of correlators. Another substantial advantage is that the theory begins with the microscopic degrees of freedom of the individual particles, which usually follow structurally simple equations of motion such as the Hamiltonian equations. Collective fields are introduced as operators extracting the desired information when needed from the microscopic degrees of freedom in the generating functional of the theory.

This paper aims at building upon this approach to find a new access to the theory of cosmological structure formation. Despite heroic efforts and ingenious new developments, it has been notoriously difficult to calculate into the nonlinear regime of second- or higher-order statistics of cosmic structures, such as the power spectrum of the cosmic density field (see \cite{1980lssu.book.....P, 1992MNRAS.254..729B, 1993MNRAS.264..375B, 1994MNRAS.267..811B, 1995A&A...296..575B, 1995ApJ...455....7M, 1997GReGr..29..733E, 2000MNRAS.318..203S, 2001A&A...379....8V, 2002PhR...372....1C, 2004A&A...421...23V, 2004MNRAS.354.1146V, 2004ApJ...612...28M, 2008PhRvD..78h3503B, 2008JCAP...10..036P, 2011JCAP...06..015A, 2012JCAP...01..019P, 2012JHEP...09..082C, 2012JCAP...12..013A, 2013PhRvD..87h3522V, 2014JCAP...05..022P, 2014JCAP...07..057C} for an inevitably eclectic list, and \cite{2002PhR...367....1B} for an extensive, impressive and complete review).

One of the cardinal difficulties in conventional approaches to cosmological structure formation, both Lagrangian and Eulerian, is that the perturbed quantities are the density and the velocity fields. With dissipationless dark matter dominating cosmic structures, however, particle trajectories can cross and form multiple streams, at which point of the evolution the description by unique and smooth density and velocity fields breaks down. This difference between fluid and particle dynamics seems to be the decisive reason for the difficulty in higher-order, standard cosmological perturbation theory.

Numerical simulations of cosmic structure formation do not encounter this problem because they follow the trajectories of large numbers of individual tracer particles. The density or other collective information is calculated when needed from the actual positions of these tracer particles.

A statistical, non-equilibrium field theory for classical particles can be seen as the analytic analog to a numerical simulation based on particles. As in simulations, following particle trajectories has the decisive advantages that the equations of motion are as simple as possible and that crossing trajectories pose no difficulty at all for the analytic treatment.

With the foundation of the theory worked out in \cite{2010PhRvE..81f1102M} and \cite{2012JSP...149..643D}, the next step towards cosmological structure formation is the definition of a suitable initial particle ensemble in phase space. Working this out is the first purpose of this paper. For rendering our discussion more accessible and self-contained, we shall begin in Sect.~\ref{sc:2} by summarising the non-equibrium field theory for classical particles. In Sect.~\ref{sc:3}, we include operators for collective fields and for the particle interactions. We ask the expert readers for patience, but we believe that this paper is more useful if it contains an outline of the theoretical foundations even though they have already been developed and described in detail elsewhere. In Sect.~\ref{sc:4}, we construct the free generating functional for particle ensembles initially correlated in phase space. Section~\ref{sc:5} discusses first-order perturbation theory in canonical particle ensembles. In Sect.~\ref{sc:6}, we derive specific expressions for low-order correlators of the density and response fields, which we then specialise in Sect.~\ref{sc:7} to derive density power spectra for cosmic structures at first order in the gravitational interaction. This section presents the first fully analytic calculation of the non-linear power spectrum of cosmic density fluctuations. Section~\ref{sc:8} summarises the paper and presents our conclusions.

Even though cosmic structure formation is our main motivation, we believe that both the approach and our central results, the generating functional for correlated classical particle ensembles and approximations to it, may be useful for other areas of statistical physics. We thus intend to lay out the formalism as generally as possible, with little or no reference to cosmology until Sect.~\ref{sc:7}. To streamline the notation, we use the abbreviations
\begin{equation}
  \int_x := \int\rmd^dx\rmd t\quad\mbox{and}\quad
  \int_k := \int\frac{\rmd^dk}{(2\pi)^d}\;,
\label{eq:01-1}
\end{equation}
where $d$ is the number of spatial dimensions.

\subsection{Summary of main concepts and results}

This paper is quite technical. To provide a compact overview, we summarise here the concepts, the approximations made and the main results. Essentially, the theory laid out here on the foundations of \cite{1973PhRvA...8..423M, 1977PhRvA..16..732F, 2013JSP...152..159D, 2012JSP...149..643D, 2011PhRvE..83d1125M, 2010PhRvE..81f1102M} begins with an initial phase-space distribution of classical particles following Hamiltonian dynamics. Like in thermodynamics, the statistical properties of this ensemble are characterised by a free generating functional (or partition function) $Z_0$ given in (\ref{eq:01-31}). This generating functional assigns a probability $P(q, p)$ for each initial phase-space position $(q, p)$ to be occupied by a particle of the ensemble. The phase-space points are then evolved forward in time. A phase factor containing the retarded Green's function of the free Hamiltonian ensures that particles move along their free, classical trajectories.

This free generating functional $Z_0$ is then extended in two ways. First, in complete analogy to quantum field theory, the particle interactions are written in form of a multiplicative, exponential operator acting on $Z_0$. Second, since the full microscopic information contained in the particle ensemble is rarely required, macroscopic or collective fields are introduced as superpositions of microscopic fields. The minimum set of collective fields consists of the number density $\rho$ of the ensemble particles and the so-called response-field $B$, with the latter describing how the evolution of the particle ensemble responds to changes in the particle coordinates by means of an interaction potential. This leads to the generating functional $Z$ in (\ref{eq:01-58}) which contains all interactions and the collective fields required.

So far, the theory is independent of the specific particle ensemble to be studied. With an eye on cosmological structure formation, the initial phase-space probability distribution $P(q,p)$ is then constructed to incorporate the appropriate auto-correlations of particle positions and momenta, and the cross-correlations between particle positions and momenta required by continuity. This results in the probability distribution (\ref{eq:01-65}), which contains a correlation operator shown in (\ref{eq:A01-44}).

Up to this point, the theory is exact. As (\ref{eq:01-65}) shows, the probability distribution contains the momentum-auto-correlation matrix of the particles in the argument of an exponential. The dependence of the auto-correlation on the particle positions needs to be integrated out, which is not generally possible analytically. Therefore, we shall expand this exponential up to the second order in the momentum auto-correlations, which is justified in cosmology because the amplitude of these correlations is low.

Similarly, the particle interactions are expressed by an exponential interaction operator. Like in quantum field theory, the series expansion of this interaction operator leads to the Feynman graphs of the theory. We shall expand the interaction operator to first order only, leaving the (tedious, but ultimately inevitable) higher-order perturbation theory to later work.

Thus, apart from the general foundations of the theory, we apply two types of approximation in this paper, viz.\ the Taylor expansions in the momentum auto-correlations to second order, and in the interaction operator to first order. In addition, but without invoking further approximations, we describe the free particle dynamics in terms of an improved version of the Zel'dovich propagation \cite{2015PhRvD..91h3524B}. Since the improved, free Zel'dovich trajectories already incorporate part of the gravitational interaction, it is interesting to see the non-linear growth of the density power spectrum possible even before explicitly including the particle interactions. This result, which is of zeroth order in the interaction operator, is found in (\ref{eq:01-182}). The contributions to the non-linear power-spectrum evolution to first order in the interaction are summarised in (\ref{eq:01-190}).

\section{Non-equilibrium statistical theory for classical fields}
\label{sc:2}

\subsection{Transition probability for classical fields}

Let $\varphi_a(t, \vec q\,)$ be a classical field with $n$ components, $1\le a\le n$, at time $t$ and position $\vec q$, living in $d$ space-time dimensions. Further, let the dynamics of this field be described by some equation of motion, written here symbolically as
\begin{equation}
  E\left(\varphi_a\right) = 0\;.
\label{eq:01-2}
\end{equation}
Any classical field $\varphi_a(t, \vec q\,)$ needs to satisfy (\ref{eq:01-2}) everywhere in the space-time domain considered.

Among all classical field configurations satisfying (\ref{eq:01-2}), one particular configuration is singled out by specifying suitable initial conditions, $\varphi_a(t_\mathrm{i}, \vec q\,) =: \varphi^\mathrm{(i)}_a(\vec q\,)$, defined at some initial instant of time $t_\mathrm{i}$ which we choose to be zero without loss of generality, $t_\mathrm{i} = 0$. The initial field configuration is mapped on a later field configuration by the classical flow $\Phi_t^\mathrm{(cl)}$,
\begin{equation}
  \varphi_a^\mathrm{(i)}(\vec q\,) \mapsto \varphi_a(t, \vec q\,) =
  \Phi^\mathrm{(cl)}_t\left(\varphi_a^\mathrm{(i)}(\vec q\,)\right)\;,\quad
  \mbox{with}\quad \Phi_0^\mathrm{(cl)} = \mathrm{id}\;.
\label{eq:01-3}
\end{equation}
Since a classical field evolves deterministically, a field configuration $\varphi_a(t, \vec q\,)$ can be reached at $t\ge0$ beginning with an initial field configuration $\varphi_a^\mathrm{(i)}(\vec q\,)$ if and only if $\varphi_a(t, \vec q\,)$ satisfies (\ref{eq:01-3}).

In analogy to quantum field theory, we aim to find the probability for the transition of an initial field configuration $\varphi_a^\mathrm{(i)}(\vec q\,)$ to a field configuration $\varphi_a(t, \vec q\,)$ at a later time $t$. This probability must be unity if and only if the evolution from $\varphi_a^\mathrm{(i)}(\vec q\,)$ to $\varphi_a(t, \vec q\,)$ follows the classical path determined by the flow $\Phi_t^\mathrm{(cl)}$.

Introducing the functional Dirac delta distribution $\delta_\mathrm{D}[\cdotp]$, we write the transition probability as a path integral,
\begin{equation}
  P\left[\varphi_a, \varphi_a^\mathrm{(i)}\right] =
  \int\D\varphi_a\,
  \delta_\mathrm{D}\left[E\left(\varphi_a\right)\right]\;.
\label{eq:01-4}
\end{equation}
The meaning of expression (\ref{eq:01-4}) is straightforward: A path integration over all possible field configurations $\varphi_a$ beginning with $\varphi_a^\mathrm{(i)}$ is being carried out, but the functional delta distribution allows only that particular path to contribute which satisfies the equation of motion.

We now introduce a conjugate field $\hat\chi_a$ to express the delta distribution by a functional Fourier transform,
\begin{equation}
  \delta_\mathrm{D}\left[E\left(\varphi_a\right)\right] =
  \int\mathcal{D}\hat\chi_a\,\exp\left\{\rmi\int_x\,
  \hat\chi_aE\left(\varphi_a\right)\right\}\;,
\label{eq:01-5}
\end{equation}
where the integration within the exponential function proceeds in general over all $d$ space-time coordinates that the fields $\varphi_a$ and $\hat\chi_a$ depend on, and a summation over $a$ is implied. Note that $\hat\chi_a$ plays the role of the ``hatted'' field $\hat\psi$ introduced by \cite{1973PhRvA...8..423M}. Even though we shall later connote operators with hats, we also add a hat here to emphasise the relation to \cite{1973PhRvA...8..423M}. The transition probability (\ref{eq:01-4}) then reads
\begin{equation}
  P\left[\varphi_a,\varphi_a^\mathrm{(i)}\right] =
  \int\D \varphi_a\int\D \hat\chi_a\,
  \exp\left\{\rmi\int_x\,\hat\chi_aE\left(\varphi_a\right)\right\}\;.
\label{eq:01-6}
\end{equation}

Identifying the integral in the exponential with the action $S$ and its integrand with the Lagrange density $\mathcal{L}$, we define
\begin{equation}
  S\left[\varphi_a, \hat\chi_a\right] := \int_x\,\mathcal{L}(x)\;,\quad
  \mathcal{L}(x) := \hat\chi_a(x)E\left(\varphi_a(x)\right)\;.
\label{eq:01-7}
\end{equation}
The functional derivative of $S$ with respect to the conjugate field $\hat\chi_a$, set to zero, reproduces the equation of motion (\ref{eq:01-2}),
\begin{equation}
  \left.
    \frac{\delta S\left[\varphi_a, \hat\chi_a\right]}{\delta\hat\chi_a(x)}
  \right\vert_{\hat\chi_a = 0} = E\left(\varphi_a(x)\right) = 0\;.
\label{eq:01-8}
\end{equation}

\subsection{Generating functional for a classical theory}

A generating functional is now readily constructed from the transition probability $P_\mathrm{fi}$. Since the path beginnung with a fixed initial field configuration is deterministic in a classical field theory, the only possible random element in such a theory is the configuration of the initial states. The configuration space to be summed or integrated over in the construction of the generating functional is thus the space of initial field configurations.

Therefore, we integrate over all possible configurations of initial states, weighted by an initial probability distribution $P_0\left[\varphi_a^\mathrm{(i)}\right]$. We shall abbreviate the path integral over the initial field configurations by
\begin{equation}
  \int\D\varphi_a^\mathrm{(i)}P_0\left[\varphi_a^\mathrm{(i)}\right] =:
\int\D\Gamma_\mathrm{i}\;.
\label{eq:01-9}
\end{equation}
Later, when we shall specify classical microscopic degrees of freedom for the fields, the initial states will be defined by a point set rather than by a set of functions. Then, the probability distribution $P_0$ for the initial conditions will be a function rather than a functional, and the path integrations over the initial states will turn into ordinary integrations.

Finally, we introduce auxiliary source fields $J_a$ for $\varphi_a$ and $K_a$ for $\hat\chi_a$ into the Lagrangian and thus arrive at the generating functional
\begin{eqnarray}
  Z[J_a, K_a] &=& \int\D\Gamma_\mathrm{i}\,
  P\left[\varphi_a,\varphi_a^\mathrm{(i)}\right]
\label{eq:01-10}
  \\&=&
  \int\D\Gamma_\mathrm{i}\int\D \varphi_a\int\D \hat\chi_a
  \exp\left[
    \rmi\int_x\left(\mathcal{L}+J_a\varphi_a+K_a\hat\chi_a\right)
  \right] \;.\nonumber
\end{eqnarray}
Functional derivatives of $Z$ with respect to the source field $J_a$, taken at $J_a = 0 = K_a$, give
\begin{eqnarray}
\label{eq:01-11}
  \left.\frac{1}{\rmi}\frac{\delta Z}{\delta J_a}\right\vert_{J = 0 = K} &=&
  \int\D\Gamma_\mathrm{i}\int\D \varphi_a\int\D\hat\chi_a\,
  \varphi_a\e^{\rmi S[\varphi_a,\hat\chi_a]} \\&=&
  \int\D\Gamma_\mathrm{i}\int\D \varphi_a\,
  \varphi_a\delta_\mathrm{D}\left[E(\varphi_a)\right] =
  \left\langle\varphi_a\right\rangle_{P_0}\;,\nonumber
\end{eqnarray}
which is the classical solution to the equation of motion, averaged over all possible initial field configurations $\varphi_a^\mathrm{(i)}$ drawn from the distribution $P_0\left[\varphi_a^\mathrm{(i)}\right]$. Field correlators are given as in quantum field theory,
\begin{eqnarray}
  &&\left\langle
    \underbrace{\varphi_{a_1}(x_1)\varphi_{a_2}(x_2)\ldots}_
    {\hbox{$m$ times}}
    \underbrace{\hat\chi_{a_{m+1}}(x_{m+1})\hat\chi_{a_{m+2}}(x_{m+2})\ldots}_
    {\hbox{$n$ times}}
  \right\rangle \nonumber\\ &=&
    \left.
    \frac{\delta}{\rmi\delta J_{a_1}(x_1)}\ldots
    \frac{\delta}{\rmi\delta K_{a_{m+n}}(x_{m+n})}
    Z[J_a, K_a]\right\vert_{J = 0 = K}\;,
\label{eq:01-12}
\end{eqnarray}
if $Z$ is normalised, $Z[0, 0] = 1$. Since we have obtained $Z$ by integration over a functional delta distribution, normalisation is ensured. Likewise, as in quantum field theory, the functional $W = \ln Z$ is the generating functional for the \emph{connected} correlators, i.e.~the cumulants.

\subsection{Generating functional for the non-interacting theory}

Suppose now that the equation of motion can be brought into the form
\begin{equation}
  E\left(\varphi_a\right) = \dot\varphi_a + E_0\left(\varphi_a\right) +
E_\mathrm{I}\left(\varphi_a\right) = 0\;,
\label{eq:01-13}
\end{equation}
where $E_0$ represents the free motion while $E_\mathrm{I}$ is due to any interaction. We can then split the action and the Lagrangian into a free part
\begin{equation}
  S_0 = \int_x\mathcal{L}_0\;,\quad
  \mathcal{L}_0 = \hat\chi_a\left[\dot\varphi_a+E_0(\varphi_a)\right]
\label{eq:01-14}
\end{equation}
and an interacting part
\begin{equation}
  S_\mathrm{I} = \int_x\mathcal{L}_\mathrm{I}\;,\quad
  \mathcal{L}_\mathrm{I} = \hat\chi_aE_\mathrm{I}(\varphi_a)\;.
\label{eq:01-15}
\end{equation}
We shall proceed with the free part first, ignoring for now any interactions between the fields $\varphi_a$ themselves or between the fields $\varphi_a$ and any external field. This will lead us to a free generating functional $Z_0[J_a,K_a]$. We shall later include the interaction part of the action in operator form, writing
\begin{equation}
  Z[J_a,K_a] = \e^{\rmi\hat S_\mathrm{I}}Z_0[J_a,K_a]
\label{eq:01-16}
\end{equation}
for the generating functional including interactions; see Sect.~III.

Restricting the action to its free part $S_0$, we obtain the generating functional for the free theory from (\ref{eq:01-10}),
\begin{eqnarray}
  \fl
  Z_0[J_a, K_a] &=& \int\D\Gamma_\mathrm{i}
  \int\D \varphi_a\int\D\hat\chi_a
  \exp\left\{\rmi\int_x\left[
    \hat\chi_a\left(\dot\varphi_a+E_0(\varphi_a)+K_a\right)+J_a\varphi_a
  \right]\right\}
  \nonumber\\ &=&
  \int\D\Gamma_\mathrm{i}\int\D\varphi_a
  \delta_\mathrm{D}\left[\dot\varphi_a+E_0(\varphi_a)+K_a\right]
  \exp\left\{\rmi\int_xJ_a\varphi_a\right\}\;.
\label{eq:01-17}
\end{eqnarray}

For any given initial field configuration $\varphi_a^\mathrm{(i)}$, the delta distribution in (\ref{eq:01-17}) singles out the solution $\bar\varphi_a(x)$ of the free equation of motion, augmented by the inhomogeneous source term $K_a$. Let $G_{ab}(x,x')$ be the propagator (Green's function) of the free equation of motion, then this solution is
\begin{equation}
  \bar\varphi_a(x) = G_{ab}(x,x_\mathrm{i})\varphi^\mathrm{(i)}_b(x_\mathrm{i})-
  \int_{x'}G_{ab}(x,x')K_b(x')\;.
\label{eq:01-18}
\end{equation}

Absorbing a constant functional determinant into the normalisation of the generating functional, we can replace the delta distribution in (\ref{eq:01-17}) by
\begin{equation}
  \delta_\mathrm{D}\left[\dot\varphi_a+E_0(\varphi_a)+K_a\right] \to
\delta_\mathrm{D}\left[\varphi_a-\bar\varphi_a\right]
\label{eq:01-19}
\end{equation}
and write the free generating functional as
\begin{equation}
  Z_0[J_a, K_a] = \int\D\Gamma_\mathrm{i}
  \exp\left\{\rmi\int_xJ_a\bar\varphi_a\right\}\;,
\label{eq:01-20}
\end{equation}
where $\varphi_a$ was replaced by the free solution $\bar\varphi_a$ from (\ref{eq:01-18}) by integrating over the delta distribution.

\section{Microscopic and collective fields}
\label{sc:3}

\subsection{Introductory remarks}

So far, the formalism chosen for classical fields is independent of the specific equations of motion and of the general properties of the fields. For the following discussion, the distinction between macroscopic and microscopic fields or degrees of freedom will be important. Instead of a macroscopic field such as an electromagnetic field, we can also use the formalism for describing the kinematics of point particles under the influence of Hamiltonian dynamics in three spatial dimensions. Then, delta distributions at the phase-space coordinates $\vec x_j^{\,\top} := (\vec q_j, \vec p_j)$ for all particles  $1\le j\le N$ replace the fields $\varphi_a$. The equations of motion of the phase-space points $\vec x_j$ are Hamilton's equations,
\begin{equation}
  \partial_t\vec x_j = \mathcal{J}\partial_j\mathcal{H}\;,
\label{eq:01-21}
\end{equation}
where $\mathcal{H}$ is the Hamiltonian, $\mathcal{J}$ is the symplectic matrix
\begin{equation}
  \mathcal{J} = \matrix{cc}{0 & \mathcal{I}_3 \\ -\mathcal{I}_3 & 0}\;,
\label{eq:01-22}
\end{equation}
and the derivative $\partial_j$ acts upon all six phase-space coordinates $x_j$ of the $j$-th particle. The matrix $\mathcal{I}_d$ is the unit matrix in $d$ dimensions. With microscopic degrees of freedom, the action $S$ in (\ref{eq:01-7}) simplifies to a time integral, as in classical mechanics. Similarly, the Green's function will then also depend on time only.

\subsection{Data structure}

For notational as well as conceptual simplicity, we follow \cite{1986ApJ...304...15B} and organise the positions $\{\vec q_j\}$ and the momenta $\{\vec p_j\}$ of $N$ microscopic particles by means of the tensor product into the phase-space coordinate tensors
\begin{equation}
  \bi q = \vec q_j\otimes\vec e_j \;,\quad \bi p = \vec p_j\otimes\vec e_j
\label{eq:01-23}
\end{equation}
where summation over repeated indices is implied, and $\vec e_j$ is the $N$-dimensional column vector whose only non-vanishing entry is 1 at component $j$. Recall that the tensor product has the convenient properties
\begin{eqnarray}
  (A\otimes B)\cdot(C\otimes D) &=& (AC)\otimes(BD)\;,\nonumber\\
  (A\otimes B)^\top &=& A^\top\otimes B^\top\;,\nonumber\\
  \mathrm{Tr}(A\otimes B) &=& \mathrm{Tr}A\cdot\mathrm{Tr}B\;.
\label{eq:01-24}
\end{eqnarray}
We further introduce the scalar product
\begin{equation}
  \langle\bi a,\bi b\rangle :=
  \left(\vec a_j\otimes\vec e_j\right)\cdot
  \left(\vec b_k\otimes\vec e_k\right) =
  \vec a_j\cdot\vec b_k\,\delta_{jk} = \vec a_j\cdot\vec b_j\;,
\label{eq:01-25}
\end{equation}
where the sum over the repeated indices is again implied. Bundling the phase-space points accordingly,
\begin{equation}
  \bi x := \vec x_j\otimes\vec e_j =
  \cvector{\vec q_j \\ \vec p_j}\otimes\vec e_j\;,
\label{eq:01-26}
\end{equation}
we can write Hamilton's equations for all $N$ particles in the compact form
\begin{equation}
  \partial_t\bi x = \left(\mathcal{J}\otimes\mathcal{I}_N\right)\cdot
  \left(\partial_j\otimes\vec e_j\right)\mathcal{H}\;.
\label{eq:01-27}
\end{equation}

Like the phase-space coordinates, we bundle the source fields $J$ and $K$ as
\begin{equation}
  \bi J = \cvector{\vec J_{q_j} \\ \vec J_{p_j}}\otimes\vec e_j\;,\quad
  \bi K = \cvector{\vec K_{q_j} \\ \vec K_{p_j}}\otimes\vec e_j\;.
\label{eq:01-28}
\end{equation}
We then need to introduce an analogous tensor product for the propagators,
\begin{equation}
  \mathcal{G} = G\otimes\mathcal{I}_N\;,
\label{eq:01-29}
\end{equation}
where $G$ is a $6\times6$ dimensional matrix describing the free propagation of an individual phase-space point.

With this notation, we can write the free solution (\ref{eq:01-18}) as
\begin{equation}
  \bar{\bi x}(t) = \mathcal{G}(t,0)\bi x^\mathrm{(i)} -
  \int_0^t\rmd t'\,\mathcal{G}(t,t')\bi K(t')
\label{eq:01-30}
\end{equation}
for all particles together, and the free generating functional assumes the form
\begin{equation}
  Z_0[\bi J, \bi K] = \int\rmd\Gamma_\mathrm{i}\,
  \exp\left\{
    \rmi\int_\mathrm{i}^\mathrm{f}\rmd t\,
    \left\langle\bi J(t),\bar{\bi x}(t)\right\rangle
  \right\}\;.
\label{eq:01-31}
\end{equation}
Note that the integral over the initial phase-space configurations is now an ordinary rather than a path integral.

\subsection{Collective fields}

If the fields $\varphi_a$ represent microscopic degrees of freedom, such as the phase-space coordinates of point particles, it will be appropriate to introduce collective fields in addition, i.e.~fields representing collective properties of the particle ensemble. Perhaps the most obvious example of such a collective field is the density $\rho(t,\vec q\,)$,
\begin{equation}
  \rho(t,\vec q\,) = \sum_{j=1}^N\delta_\mathrm{D}\left(\vec q-\vec
q_j(t)\right)\;,
\label{eq:01-32}
\end{equation}
here assumed to be composed of $N$ point-particle contributions.

The potential $V(t,\vec q\,)$ experienced by any particle at time $t$ and at position $\vec q$ is the sum over all point-particle potentials $v$,
\begin{equation}
  V(t,\vec q\,) = \sum_{j=1}^Nv\left(\vec q-\vec q_j(t)\right)\;,
\label{eq:01-33}
\end{equation}
which we can re-write in terms of an integral over the density (\ref{eq:01-32}),
\begin{equation}
  V(t,\vec q\,) = \int\rmd^3y\,v\left(\vec q-\vec y\,\right)\,
  \sum_{j=1}^N\delta_\mathrm{D}\left(\vec y-\vec q_j(t)\right) =
  \int\rmd^3y\,v\left(\vec q-\vec y\,\right)\,\rho(t,\vec y\,)\;.
\label{eq:01-34}
\end{equation}

According to Hamilton's equations, and given the interaction potential $V(t, \vec q\,)$, the interaction contribution to the equations of motion of the particle ensemble is
\begin{equation}
  \bi E_\mathrm{I}(\bi q) = -\left(\mathcal{J}\otimes\mathcal{I}_N\right)
  \left(\partial_j\otimes\vec e_j\right)V(t,\vec q\,) =
  \cvector{0 \\ \partial_{q_j}}V(t,\vec q\,)\otimes\vec e_j\;,
\label{eq:01-35}
\end{equation}
where $\partial_{q_j}$ is the spatial derivative of $V(t,\vec q\,)$ taken at
$\vec q = \vec q_j$,
\begin{eqnarray}
  \partial_{q_j}V(t,\vec q\,) &= \left.\partial_qV(t,\vec q\,)\right\vert_{\vec
q=\vec q_j} =
  \int\rmd^3q\,\delta_\mathrm{D}\left(\vec q-\vec q_j\right)\partial_qV(t,\vec
q\,) \nonumber\\ &=
  -\int\rmd^3q\,\left[\partial_q\delta_\mathrm{D}\left(\vec q-\vec
q_j\right)\right]V(t,\vec q\,)\;,
\label{eq:01-36}
\end{eqnarray}
where the last step was taken by partial integration to remove the gradient from the potential for later convenience. With this result and with the conjugate field $\hat{\bchi} := \hat\chi_j\otimes\vec e_j$, we can thus write the interaction contribution to the Lagrange density $\mathcal{L}_\mathrm{I}$ from (\ref{eq:01-15}) as
\begin{equation}
  \mathcal{L}_\mathrm{I} = \left\langle\hat{\bchi},\bi E_\mathrm{I}(\bi q)\right\rangle =
  -\int\rmd^3q\,\left[\sum_{j=1}^N
    \hat\chi_{p_j}\cdot\partial_q
    \delta_\mathrm{D}\left(\vec q-\vec q_j(t)\right)
  \right]V(t,\vec q\,)\;.
\label{eq:01-37}
\end{equation}
The term in brackets defines the \emph{response field}
\begin{equation}
  B(t,\vec q\,) := \sum_{j=1}^N
  \hat\chi_{p_j}\cdot\partial_q
  \delta_\mathrm{D}\left(\vec q-\vec q_j(t)\right)
\label{eq:01-38}
\end{equation}
in terms of the conjugate ``hatted'' field $\hat{\bchi}$. Introducing this into (\ref{eq:01-37}), we can write the interaction part of the Lagrange density as
\begin{eqnarray}
  \mathcal{L}_\mathrm{I} = -\int\rmd^3q\,B(t,\vec q\,)V(t,\vec q\,) =
  -\int\rmd^3q\int\rmd^3y\,
    B(t,\vec q\,)v\left(\vec q-\vec y\,\right)\rho(t,\vec y\,)\;.
\label{eq:01-39}
\end{eqnarray}
Expressing the response field $B$, the potential $v$ and the density $\rho$ by their Fourier transforms, we can re-write the interaction Lagrangian as
\begin{equation}
  \mathcal{L}_\mathrm{I} = -\int_k
  B\left(t,-\vec k\,\right)v\left(\vec k\,\right)\rho\left(t,\vec k\,\right)\;,
\label{eq:01-40}
\end{equation}
where we have assumed that the potential $v$ in (\ref{eq:01-39}) is translation invariant and thus depends on the difference $\vec q-\vec y$ only.

We now combine the two collective fields $\rho(t,\vec k\,)$ and $B(t,\vec k\,)$ into the field doublet
\begin{equation}
  \Phi\left(t,\vec k\,\right) :=
  \cvector{\rho\left(t,\vec k\,\right)\\ B\left(t,\vec k\,\right)}
\label{eq:01-41}
\end{equation}
and write the interaction part of the action in the compact form
\begin{equation}
  S_\mathrm{I}(t) = \frac{1}{2}\int\rmd 1\int\rmd 2\,
  \Phi^\top(-1)\sigma(12)\Phi(2)\;,
\label{eq:01-42}
\end{equation}
where the conventional abbreviations $\rmd1 = \rmd t_1\rmd^3k_1$ and $\Phi(1) = \Phi(t_1,\vec k_1)$ are being used. We further write $\Phi(-1) = \Phi(t_1,-\vec k_1)$. The quantity $\sigma(12)$ is defined to be the \emph{interaction matrix}
\begin{equation}
  \sigma(12) = -v(1)
  \delta_\mathrm{D}(1-2)\matrix{cc}{0&1\\1&0}\;,
\label{eq:01-43}
\end{equation}
where the delta distribution $\delta_\mathrm{D}(1-2)$ enters because of the spatial translation invariance of the potential $v$ and the assumed instantaneous interaction. The doublet $\Phi = (\rho, B)$ of collective fields must be paired with a doublet $H = (H_\rho, H_B)$ of conjugate source fields in the Lagrangian and in the action. We thus extend the free part of the action as
\begin{equation}
  S_0 \to S_0 + H\cdot\Phi\;,
\label{eq:01-44}
\end{equation}
where the product is understood as an implicit sum over the collective-field
indices and integral over the space-time coordinates,
\begin{equation}
  H\cdot\Phi= \sum_a\int\rmd1\,H_a(1)\Phi_a(1)\;.
\label{eq:01-45}
\end{equation}

\subsection{Operator expressions for the collective fields}

The collective fields $\Phi$ typically contain the field variables $\bi x$ or $\hat{\bchi}$. These are obtained from the free functional $Z_0[\bi J,\bi K]$ by functional derivatives with respect to the sources $\bi J$ or $\bi K$,
\begin{equation}
  \bi x(t) \to \frac{\delta}{\rmi\delta\bi J(t)}\;,\quad
  \hat{\bchi}(t) \to \frac{\delta}{\rmi\delta\bi K(t)}\;.
\label{eq:01-46}
\end{equation}
For introducing the values of $\bi x$ and $\hat{\bchi}$ into $\Phi$, we replace $\Phi$ by an operator $\hat\Phi$ acting on the free functional $Z_0[\bi J,\bi K]$, with all occurrences of $\bi x$ and $\hat{\bchi}$ replaced by functional derivatives according to (\ref{eq:01-46}). Then, the free generating functional including the collective fields is expressed by
\begin{equation}
  Z_0[H,\bi J,\bi K] =
  \exp\left(\rmi H\cdot\hat\Phi\right)Z_0[\bi J,\bi K]\;.
\label{eq:01-47}
\end{equation}
The minimum set of collective fields that we require are the density $\rho$ and the response field $B$. We shall now construct their operator expressions.

The density $\rho$ is assumed to be composed of delta contributions, see (\ref{eq:01-32}). In Fourier space, the one-particle contribution of particle $j$ to the density at the space-time position $1 = (t_1,\vec k_1)$ is
\begin{equation}
  \rho_j(1) = \exp\left(-\rmi\vec k_1\cdot\vec q_j(t_1)\right)\;,
\label{eq:01-48}
\end{equation}
where $\vec q_j(t_1)$ is the position of particle number $j$ in configuration space at time $t_1$. In this expression for the density, we replace the particle position $\vec q_j$ by a functional derivative with respect to $\vec J_{q_j}(1)$, obtaining the one-particle density operator
\begin{equation}
  \hat\Phi_{\rho_j}(1) = \exp\left(-\rmi\vec k_1^{\,\top}\cdot
  \frac{\delta}{\rmi\delta\vec J_{q_j}(1)}\right)\;.
\label{eq:01-49}
\end{equation}

The action of the density operator (\ref{eq:01-49}) on the free generating
functional (\ref{eq:01-31}) becomes clear by expanding the exponential into a
series. This immediately leads to
\begin{eqnarray}
  \hat\Phi_{\rho_j}(1)Z_0[\bi J,\bi K] = Z_0[\bi J+\bi L_j(1),\bi K]\;,
\label{eq:01-50}
\end{eqnarray}
where the tensor $\bi L_j(1)$ is defined by
\begin{equation}
  \bi L_j(1) := -\vec k_1\cdot\frac{\delta\bi J(t)}{\delta\vec J_{q_j}(1)} =
  -\delta_\mathrm{D}\left(t-t_1\right)\cvector{\vec k_1\\ 0}\otimes\vec e_j\;.
\label{eq:01-51}
\end{equation}
Thus, the application of the density operator $\hat\Phi_{\rho_j}(1)$ amounts to a shift of the source field $\bi J$ in the free generating functional by the tensor $\bi L_j(1)$.

According to (\ref{eq:01-38}), the one-particle contribution $B_j(1)$ of particle $j$ to the collective field $B(1)$ is determined by the gradient of the density,
\begin{equation}
  B_j(1) = \hat\chi_{p_j}\cdot
  \partial_{q_1}\delta_\mathrm{D}\left(\vec q_1-\vec q_j(t_1)\right)\;.
\label{eq:01-52}
\end{equation}
Taken into Fourier space, the one-particle response-field operator thus turns into
\begin{equation}
  \hat\Phi_{B_j}(1) =
  \left(\rmi\vec k_1^{\,\top}\cdot
  \frac{\delta}{\rmi\delta\vec K_{p_j}(1)}\right)\hat\Phi_{\rho_j}(1) =:
  \hat b_j(1)\hat\Phi_{\rho_j}(1)\;.
\label{eq:01-53}
\end{equation}
Since $\hat\Phi_{\rho_j}$ involves functional derivatives with respect to $\vec J_{q_j}$ while the response-field operator takes functional derivatives with respect to $\vec K_{p_j}$, the relevant functional derivatives commute. We can thus reorder the operators and apply all required response-field operators after all density operators.

\subsection{Operator expression for the interaction part of the action}

We have seen in (\ref{eq:01-42}) and (\ref{eq:01-43}) that the interaction between the particles can be included by adding the expression
\begin{eqnarray}
  S_\mathrm{I} =
  -\int\rmd 1\int\rmd 2\,B(-1)\,v(1)\delta_\mathrm{D}(1-2)\,\rho(2) =
  -\int\rmd 1\,B(-1)v(1)\rho(1)
\label{eq:01-54}
\end{eqnarray}
to the free action. By means of the operator expressions (\ref{eq:01-49}) and (\ref{eq:01-53}) for the two collective fields $\rho$ and $B$, we can write this action contribution in the operator form
\begin{equation}
  \hat S_\mathrm{I} = -\int\rmd 1\,\hat\Phi_B(-1)\,v(1)\hat\Phi_\rho(1)\;,
\label{eq:01-55}
\end{equation}
where the operators $\hat\Phi_\rho$ and $\hat\Phi_B$ will now be responsible for acquiring the respective collective-field values from the free generating functional $Z_0[H,\bi J,\bi K]$.

Written in the form (\ref{eq:01-47}), this free generating functional contains the collective fields in operator form already, paired with their conjugate source fields $H_\rho$ and $H_B$. The collective-field operators themselves are thus obtained from $Z_0[H,\bi J,\bi K]$ by functional derivatives with respect to the conjugate $H$ fields,
\begin{equation}
  \hat\Phi_\rho(1) \to \frac{\delta}{\rmi\delta H_\rho(1)}\;,\quad
  \hat\Phi_B(-1) \to \frac{\delta}{\rmi\delta H_B(-1)}
\label{eq:01-56}
\end{equation}
applied to $Z_0[H,\bi J,\bi K]$. Therefore, the interaction part of the action is
\begin{equation}
  \hat S_\mathrm{I} = -\int\rmd 1\,
  \frac{\delta}{\rmi\delta H_B(-1)}\,v(1)\,
  \frac{\delta}{\rmi\delta H_\rho(1)}\;,
\label{eq:01-57}
\end{equation}
allowing us to express the complete generating functional as
\begin{equation}
  Z[H,\bi J,\bi K] = \e^{\rmi\hat S_\mathrm{I}}\e^{\rmi H\cdot\hat\Phi}
  Z_0[\bi J,\bi K]\;,
\label{eq:01-58}
\end{equation}
understanding that $H\cdot\hat\Phi$ abbreviates the expression
\begin{equation}
  H\cdot\hat\Phi = \sum_{a=\rho, B}\int\rmd 1'\,H_a(1')\hat\Phi_a(1')
\label{eq:01-59}
\end{equation}
as noted in (\ref{eq:01-45}) before in more general form.

\section{Generating functional for correlated initial conditions}
\label{sc:4}

Having developed the general formalism for a set of Hamiltonian point particles, the final step to be taken towards defining the generating functional (\ref{eq:01-58}) completely concerns the initial phase-space distribution. In this Section, we consider points in a domain of phase space at an initial time only. For convenience, we shall drop the superscript (i) on any quantity here, understanding that all quantities are to be taken at the initial time throughout this Section.

\subsection{Initial phase-space distribution}
\label{ssc:4.1}

So far, the construction of the non-equilibrium field theory for classical particles has been completely general, with the one exception that we required the microscopic degrees of freedom to follow the Hamiltonian equations of motion. To specify the generating functional completely, we now have to define the initial phase-space measure
\begin{equation}
  \rmd\Gamma = P\left(\bi q, \bi p\right)\rmd\bi q\rmd\bi p\;;
\label{eq:01-60}
\end{equation}
that is, we have to construct the probability distribution $P(\bi q, \bi p)$ for initial particle positions in phase space.

Having cosmological structure formation in mind, we need the particles to be spatially correlated such that their number density is a homogeneous and isotropic Gaussian random field. By continuity, spatial correlations imply correlations also in momentum space as well as cross-correlations between spatial and momentum coordinates. Our main goal here is thus to derive the probability distribution for the initial phase-space coordinates under these requirements.

A central (and, as we shall see, a sufficient) quantity characterising all required correlations is the power spectrum of density fluctuations. Calling the \emph{number} density of particles $\rho$ and its mean $\bar\rho$, the density contrast is
\begin{equation}
  \delta := \frac{\rho-\bar\rho}{\bar\rho}\;,
\label{eq:01-61}
\end{equation}
and its power spectrum $P_\delta(k)$ is defined as
\begin{equation}
  \left\langle
    \delta\left(\vec k\,\right)\delta\left(\vec k'\,\right)
  \right\rangle =
  (2\pi)^3\delta_\mathrm{D}\left(\vec k+\vec k'\right)P_\delta(k)\;,
\label{eq:01-62}
\end{equation}
where the density contrast $\delta(\vec k)$ as a function of the wave vector $\vec k$ implicitly denotes the Fourier transform. The delta distribution ensures translation invariance and thus the statistical homogeneity of the density contrast. If it is statistically isotropic as well, the power spectrum depends on the wave number $k$ only and not on the direction of the wave vector $\vec k$.

The cosmological motivation aside for now, we are thus aiming to derive the probability distribution for correlated phase-space points drawn from a statistically homogeneous and isotropic Gaussian random field characterised by the power spectrum of the density fluctuations. Such initial conditions may be interesting far beyond cosmology.

For clarity of the discussion in the main part of this paper, we shall develop the probability distribution $P(\bi q, \bi p)$ in \ref{app:A}. The central variables in the derivation of this probability distribution will be the values $\delta_j$ of the density contrast and $\vec p_j$ of the momentum at the positions of all particles $j$. We organise these variables at all $N$ particle positions into a data tensor
\begin{equation}
  \bi d := \cvector{\delta\\ \vec p}_j\otimes\vec e_j
\label{eq:01-63}
\end{equation}
by means of the tensor product with the vectors $\vec e_j$ defined in (\ref{eq:01-23}). A major intermediate result will be the covariance matrix
\begin{equation}
  \bar C := \left\langle\bi d\otimes\bi d\right\rangle
\label{eq:01-64}
\end{equation}
of this data tensor, which contains the density-contrast and momentum auto-correlations $\langle\delta_j\delta_k\rangle$ and $\langle\vec p_j\otimes\vec p_k\rangle$, respectively, and the density-momentum cross-correlation $\langle\delta_j\vec p_k\rangle$. The entries of the covariance matrix are detailed in \ref{app:A}. As derived there, the initial phase-space probability distribution is
\begin{equation}
  P\left(\bi q, \bi p\right) =
  \frac{V^{-N}}{\sqrt{(2\pi)^{3N}\det\bar C_{pp}}}\,
  \mathcal{C}\left(\bi p\right)
  \exp\left(-\frac{1}{2}\bi p^\top\bar C_{pp}^{-1}\bi p\right)\;,
\label{eq:01-65}
\end{equation}
with $\bi q$ and $\bi p$ as defined in (\ref{eq:01-23}). The correlation operator $\mathcal{C}(\bi p)$ appearing here is given in (\ref{eq:A01-44}). It contains the correlation matrices $C_{\delta\delta}$ and $C_{\delta p}$ introduced above and defined in (\ref{eq:A01-32}).

If the correlations $C_{\delta\delta}$ and $C_{\delta p}$ are weak, as we can expect them to be early in time, the
probability distribution (\ref{eq:01-65}) can be approximated by
\begin{eqnarray}
  \fl
  P(\bi q, \bi p) \approx
  \frac{V^{-N}}{\sqrt{(2\pi)^{3N}\det\bar C_{pp}}}
  \exp\left(-\frac{1}{2}\bi p^\top\bar C_{pp}^{-1}\bi p\right)
  \left(
    1+\sum_{j=1}^NM_{\delta_jp_k}\vec p_k+
    \frac{1}{2}\sum_{j\ne k}C_{\delta_j\delta_k}
  \right)
\label{eq:01-66}
\end{eqnarray}
with
\begin{equation}
  M_{\delta_jp_k} := C_{\delta_jp_a}^\top\bar C_{p_ap_k}^{-1}\;.
\label{eq:01-67}
\end{equation}

\subsection{General expressions for density correlators}
\label{ssc:4.2}

We are aiming at calculating $m$-point correlators of collective fields, such as the density and response fields. Equation (\ref{eq:01-53}) shows that the response-field operator contains the density operator as a factor. Thus, for an $m$-point correlator, $m$ density operators will have to be applied to the free generating functional first. Since no further derivatives with respect to $\bi J$ will be required afterwards, the source field $\bi J$ can then be set to zero.

The operator for the density contributions by $N$ particles is the sum over the one-particle density operators,
\begin{equation}
  \hat\Phi_\rho(1) = \sum_{j=1}^N\hat\Phi_{\rho_j}(1)\;.
\label{eq:01-68}
\end{equation}

As illustrated in (\ref{eq:01-51}) for a single one-particle density operator, the result of applying $m$ one-particle density operators to the free generating functional is
\begin{equation}
  \left.
    \hat\Phi_{\rho_{j_m}}(m)\cdots\hat\Phi_{\rho_{j_1}}(1)Z_0[\bi J,\bi K]
  \right\vert_{\bi J = 0} = Z_0[\bi L,\bi K]
\label{eq:01-69}
\end{equation}
with
\begin{equation}
  \bi L = -\sum_{s=1}^m\delta_\mathrm{D}(t-t_s)\cvector{\vec k_s\\0}
  \otimes\vec e_{j_s}\;.
\label{eq:01-70}
\end{equation}
Following (\ref{eq:01-53}), a single one-particle response-field operator $\hat b_{j_l}(l)$ applied subsequently then leads to
\begin{eqnarray}
  \left.\hat b_{j_l}(l)Z_0[\bi L,\bi K]\right\vert_{\bi K=0} &=&
  \left.\left(
    \rmi\vec k_l^{\,\top}\cdot\frac{\delta}{\rmi\delta\vec K_{p_{j_l}}(l)}
  \right)Z_0[\bi L,\bi K]\right\vert_{\bi K=0} \nonumber\\ &=&
  \left(
    -\rmi\sum_{s=1}^mg_{qp}(t_s,t_l)\vec k_l\cdot\vec k_s\delta_{j_sj_l}
  \right)
  Z_0[\bi L,0]\;.
\label{eq:01-71}
\end{eqnarray}

According to (\ref{eq:01-68}) and (\ref{eq:01-69}), an $m$-point density correlator $G_{\rho\ldots\rho}(1\ldots m)$ is found by summing over all particle indices,
\begin{eqnarray}
  G_{\rho\ldots\rho}(1\ldots m) = \sum_{j_1\ldots j_m=1}^N\left.
  \hat\Phi_{\rho_{j_1}}\cdots\hat\Phi_{\rho_{j_m}}Z_0[\bi J,\bi K]
  \right\vert_{\bi J = 0 = \bi K} =
  \sum_{j_1\ldots j_m=1}^NZ_0[\bi L,0]\;.
\label{eq:01-72}
\end{eqnarray}
This shows that all we have to evaluate for $m$-point correlators of the density and response fields is the free generating functional taken at $\bi J = \bi L$ and $\bi K = 0$,
\begin{equation}
  Z_0[\bi L,0] = \int\rmd\Gamma\exp\left\{
    \rmi\int\rmd t\,\left\langle
      \bi L,\bar{\bi x}_0
    \right\rangle
  \right\}\;,
\label{eq:01-73}
\end{equation}
where the free phase-space trajectory
\begin{equation}
  \bar{\bi x}_0(t) = \mathcal{G}(t,0)\bi x^\mathrm{(i)}
\label{eq:01-74}
\end{equation}
appears because the term containing the source $\bi K$ disappears. With (\ref{eq:01-70}), the phase in the exponential in (\ref{eq:01-73}) is
\begin{eqnarray}
  \int\rmd t\,\left\langle\bi L,\bar{\bi x}_0\right\rangle =
  -\sum_{s=1}^m\left(
    \vec k_s\cdot\vec q_{j_s}^\mathrm{\,(i)}+
    \vec T_s\cdot\vec p_{j_s}^\mathrm{\,(i)}
  \right)\;,
\label{eq:01-75}
\end{eqnarray}
where $\vec T_s := g_{qp}(t_s,0)\vec k_s$ was defined for brevity. If we introduce the tensors
\begin{equation}
  \bi L_q := -\sum_{s=1}^m\vec k_s\otimes\vec e_{j_s}\quad\mbox{and}\quad
  \bi L_p := -\sum_{s=1}^m\vec T_s\otimes\vec e_{j_s}\;,
\label{eq:01-76}
\end{equation}
we can briefly write
\begin{equation}
  Z_0[\bi L,0] = \int\rmd\Gamma_\mathrm{i}\e^{
    \rmi\left\langle\bi L_q,\bi q\right\rangle+
    \rmi\left\langle\bi L_p,\bi p\right\rangle
  }
\label{eq:01-77}
\end{equation}
for the free generating functional evaluated at $\bi J = \bi L$ and $\bi K = 0$.

The shift tensors $\bi L_{q,p}$ have non-vanishing components only for the particles specified by the indices $j_s$ set by the one-particle density operators applied to the free generating functional. For any shift tensor specified by a complete set of $m$ particle indices $j_1\ldots j_m$, we write briefly
\begin{equation}
  Z_0[\bi L,0] = G_{\rho_{j_1}\ldots\rho_{j_m}}\;,\quad
  G_{\rho\ldots\rho}(1\ldots m) =
  \sum_{j_1\ldots j_m=1}^NG_{\rho_{j_1}\ldots\rho_{j_m}}\;.
\label{eq:01-78}
\end{equation}
The integral over the initial phase-space configuration remaining in the free generating functional $Z_0[\bi L,0]$ still has to be carried out.

\subsection{Integration over the initial phase-space distribution}
\label{ssc:4.3}

Inserting (\ref{eq:01-66}) into (\ref{eq:01-77}), we first find
\begin{eqnarray}
  Z_0[\bi J,0] &= V^{-N}\int\!\!\int
  \frac{\rmd\bi q\rmd\bi p}{\sqrt{(2\pi)^{3N}\det\bar C_{pp}}}
  \exp\left(-\frac{1}{2}\bi p^\top\bar C_{pp}^{-1}\bi p\right)
  \nonumber\\ &\cdot
  \left(
    1+\sum_{j=1}^NM_{\delta_jp_k}\vec p_k+
    \frac{1}{2}\sum_{j\ne k}C_{\delta_j\delta_k}
  \right)
  \e^
   {\rmi\left\langle\bi L_q,\bi q\right\rangle+
    \rmi\left\langle\bi L_p,\bi p\right\rangle}\;.
\label{eq:01-79}
\end{eqnarray}
Since the first and the third term in parentheses in the second line do not depend on the momenta $\bi p$, they can easily be integrated over $\bi p$ using
\begin{equation}
  \int\frac{\rmd\bi p}{\sqrt{(2\pi)^{3N}\det\bar C_{pp}}}
  \exp\left(-\frac{1}{2}\bi p^\top\bar C_{pp}^{-1}\bi p\right)
  \e^{\rmi\left\langle\bi L_p,\bi p\right\rangle} = \e^{-\bar Q/2}
\label{eq:01-80}
\end{equation}
with the quadratic form
\begin{equation}
  \bar Q := \bi L_p^\top\bar C_{pp}\bi L_p\;.
\label{eq:01-81}
\end{equation}
The second term in parentheses can be integrated after pulling a factor $\bi p$ down by applying the derivative $-\rmi\partial/\partial\bi L_p$ to the phase factor $\e^{\rmi\left\langle\bi L_p,\bi p\right\rangle}$. Using further that $M_{\delta p} = C_{\delta p}\bar C_{pp}^{-1}$ as defined in (\ref{eq:01-67}), this leads to
\begin{eqnarray}
  \fl
  \int\!\!\frac{\rmd\bi p}{\sqrt{(2\pi)^{3N}\det\bar C_{pp}}}
  \exp\left(-\frac{1}{2}\bi p^\top\bar C_{pp}^{-1}\bi p\right)\left(
    \sum_{j=1}^NM_{\delta_jp_k}\vec p_k
  \right)
  \e^{\rmi\left\langle\bi L_p,\bi p\right\rangle} =
  \rmi\left(
    \sum_{j=1}^NC_{\delta_jp_k}\vec L_{p_k}
  \right)\e^{-\bar Q/2}\;.
\label{eq:01-82}
\end{eqnarray}

Combining the results (\ref{eq:01-79}), (\ref{eq:01-80}) and (\ref{eq:01-82}), we find
\begin{eqnarray}
  \fl
  Z_0[\bi L,0] = V^{-N}\e^{-Q_D/2}
  \int\rmd\bi q\,\left(
    1+\rmi\sum_jC_{\delta_jp_k}L_{p_k}+
    \frac{1}{2}\sum_{j\ne k}C_{\delta_j\delta_k}
  \right)
  \e^{-Q/2+\rmi\left\langle\bi L_q,\bi q\right\rangle}\;,
\label{eq:01-83}
\end{eqnarray}
where we have used (\ref{eq:A01-32}) to split up the quadratic form $\bar Q$ into $\bar Q = Q_D + Q$, with
\begin{equation}
  Q_D := \frac{\sigma_1^2}{3}\left\langle\bi L_p,\bi L_p\right\rangle
  \quad\mbox{and}\quad
  Q := \bi L_p^\top\left[
    C_{p_jp_k}\otimes E_{jk}
  \right]\bi L_p\;.
\label{eq:01-84}
\end{equation}

For later convenience, we introduce one-particle shift vectors $\vec L_{q_j}$ and $\vec L_{p_j}$ by the projection
\begin{equation}
  \vec L_{q_j} := \left(
    \mathcal{I}_3\otimes\vec e_j
  \right)\bi L_q\;,\quad
  \vec L_{p_j} := \left(
    \mathcal{I}_3\otimes\vec e_j
  \right)\bi L_p\;,
\label{eq:01-85}
\end{equation}
which, with (\ref{eq:01-76}), turn out to be
\begin{equation}
  \vec L_{q_j} = -\sum_{s=1}^m\vec k_s\delta_{jj_s}\;,\quad
  \vec L_{p_j} = -\sum_{s=1}^m\vec T_s\delta_{jj_s}\;.
\label{eq:01-86}
\end{equation}
In terms of $\vec L_{p_j}$, we can write the quadratic forms in (\ref{eq:01-84}) as
\begin{eqnarray}
  Q_D = \frac{\sigma_1^2}{3}\sum_{j=1}^N\vec L_{p_j}^{\,2}
  \quad\mbox{and}\quad
  Q = \vec L_{p_j}^{\,\top}C_{p_jp_k}\vec L_{p_k}\;.
\label{eq:01-87}
\end{eqnarray}

\subsection{Damping}
\label{ssc:4.4}

The exponential prefactor $\exp(-Q_D/2)$ with the quadratic form $Q_D$ from (\ref{eq:01-83}) or (\ref{eq:01-87}) appearing in the free generating functional (\ref{eq:01-83}) requires a separate consideration. As the derivation of $Z_0[\bi J,\bi K]$ shows, it originates from the initial one-point momentum variance and thus arises from the free streaming of the particles with the initial root mean-square velocity quantified by $\sigma_1$. In absence of momentum correlations, it would lead to a Maxwellian or thermal velocity distribution of the particles.

In the cold-dark matter cosmogony, free streaming is suppressed by the long-ranged gravitational interaction between the massive particles. In the free generating functional, gravitational interaction is not included by definition. Later, we shall introduce gravitational interaction between the particles in a perturbative approach. As long as we neglect gravitational two-particle interaction, the damping expressed by $\exp(-Q_D/2)$ will be unrealistic for cold dark matter because it is counteracted by the gravitational interaction.

The effect of the damping term depends on its relation to the quadratic form $Q$, also defined in (\ref{eq:01-84}), which contains the initial momentum correlations between different particles. As we shall see later, these initial momentum correlations will be mainly responsible for the growth of structures. Realising that $Q$ will commonly be a small number, we shall approximate
\begin{equation}
  \e^{-Q/2} \approx 1-\frac{Q}{2}+\frac{Q^2}{8}\;,
\label{eq:01-88}
\end{equation}
i.e.\ we shall expand in powers of the initial momentum correlations. For appropriately suppressing the damping term relative to the growth of structures, we shall approximate the damping term $\exp(-Q_D/2)$ consistently at one order less than the term (\ref{eq:01-88}). This implies that damping will only be included at loop order, but not at tree order. While this may appear arbitrary here, we shall show in a follow-up paper how the damping term is counteracted when the complete hierarchy of momentum auto-correlations is taken into account. Since this calculation is quite involved, we postpone it here.

Thus, when we derive results from the free generating functional restricted to linear initial momentum correlations, we shall ignore the damping term completely, approximating
\begin{equation}
  \exp\left(-Q_D/2\right) \approx 1\;.
\label{eq:01-89}
\end{equation}
At the next higher order of the
initial momentum correlations, we shall include the
damping term at linear order, approximating
\begin{equation}
  \exp\left(-Q_D/2\right) \approx 1-\frac{Q_D}{2} \approx
  \left(1+\frac{Q_D}{2}\right)^{-1}\;,
\label{eq:01-90}
\end{equation}
where the second approximation is advantageous because it remains positive definite.

\subsection{One-point functional and normalisation}
\label{ssc:4.5}

If we consider deriving free one-point ``correlators'' of a collective field, e.g.\ of the density, from the generating functional, we can ignore all correlation terms because they appear only if two or more points are involved. Then, $Q = 0$, and the generating functional (\ref{eq:01-83}) shrinks to
\begin{equation}
  Z_0[\bi L,0] = V^{-N}\e^{-Q_D/2}\int\rmd\bi q\,
  \e^{\rmi\left\langle\bi L_q,\bi q\right\rangle}\;.
\label{eq:01-91}
\end{equation}
If a single point is involved, $\bi L_q$ will have a single non-vanishing component, which we can without loss of generality label with the index $j = 1$. Then,
\begin{equation}
  \bi L_{q} = \vec L_{q_1}\otimes\vec e_1\;,\quad
  \left\langle\bi L_q,\bi q\right\rangle =
  -\vec k_1\cdot\vec q_1\;.
\label{eq:01-92}
\end{equation}
This, inserted into (\ref{eq:01-91}), gives
\begin{equation}
  Z_0[\bi L,0] = NV^{-1}(2\pi)^3\delta_\mathrm{D}\left(\vec L_{q_1}\right)
\label{eq:01-93}
\end{equation}
because the delta distribution further ensures that $\vec L_{q_1} = 0$ and thus also $Q_D = 0$. The factor $N$ in (\ref{eq:01-93}) takes into account that there are $N$ possibilities to select a particle from the ensemble. Since the remaining delta distribution is the Fourier transform of unity,
\begin{equation}
  (2\pi)^3\delta_\mathrm{D}\left(\vec L_{q_1}\right) = \hat 1\;,
\label{eq:01-94}
\end{equation}
we see that (\ref{eq:01-93}) simply reproduces the mean particle density, as it should.

\subsection{Low-order approximations}
\label{ssc:4.6}

We can now Taylor-expand the factor $\e^{-Q/2}$ in (\ref{eq:01-83}) in powers of $Q$, for example to first or second order. This gives the two contributions
\begin{equation}
  Z_0[\bi L,0] \approx Z_0^{(1)}[\bi L,0]+Z_0^{(2)}[\bi L,0]
\label{eq:01-95}
\end{equation}
with
\begin{eqnarray}
  Z_0^{(1)}[\bi L,0] &:= V^{-N}\e^{-Q_D/2} %\nonumber\\ &\cdot
  \int\rmd\bi q\,\left(
    1-\frac{Q}{2}+\rmi\sum_jC_{\delta_jp_k}\vec L_{p_k}+
    \frac{1}{2}\sum_{j\ne k}C_{\delta_j\delta_k}
  \right)
  \e^{\rmi\left\langle\bi L_q,\bi q\right\rangle}\;, \nonumber\\
  Z_0^{(2)}[\bi L,0] &:= \frac{V^{-N}}{8}\e^{-Q_D/2}
  \int\rmd\bi q\,Q^2\,
  \e^{\rmi\left\langle\bi L_q,\bi q\right\rangle}\;.
\label{eq:01-96}
\end{eqnarray}
We only consider higher-order momentum correlations here and ignore cross terms of the form $QC_{\delta_j\delta_k}$ or $QC_{\delta_jp_k}\vec L_{p_k}$ because terms containing momentum correlations dominate at late times due to the time dependence of the momentum propagator $g_{qp}(\tau, 0)$. We shall now evaluate these expressions in detail.

The first integral in $Z_0^{(1)}[\bi L,0]$,
\begin{equation}
  \mathcal{N} := \int\rmd\bi q\,\e^{\rmi\left\langle\bi L_q,\bi q\right\rangle}\;,
\label{eq:01-97}
\end{equation}
has a particular meaning. It simply returns a delta distribution or products thereof, depending on $\bi L_q$. Since it does not incorporate any correlations, it can only represent shot-noise terms or powers of the mean particle density $\bar\rho$. For a general discussion of shot-noise terms, see Sect.~\ref{sc:5} below. Besides, all other terms involve double sums over $j,k = 1\ldots N$ and thus dominate $Z_0^{(1)}[\bi L,0]$ by far. We shall thus ignore $\mathcal{N}$ in $Z_0^{(1)}[\bi L,0]$.

In the terms remaining in (\ref{eq:01-96}), we pull the sums out of the integrals and write
\begin{eqnarray}
  Z_0^{(1)}[\bi L,0] &= V^{-N}\e^{-Q_D/2}\sum_{j\ne k=1}^N
  Z_{jk}^{(1)} \;,\nonumber\\
  Z_0^{(2)}[\bi L,0] &= \frac{V^{-N}}{8}
  \e^{-Q_D/2}\sum_{j\ne k, l\ne m=1}^NZ_{jklm}^{(2)}
\label{eq:01-98}
\end{eqnarray}
with
\begin{equation}
  Z_{jk}^{(1)} := \int\rmd\bi q\,
  \e^{\rmi\left\langle\bi L_q,\bi q\right\rangle}
  \left\{\frac{1}{2}\left(
    C_{\delta_j\delta_k}-
    \vec L_{p_j}^\top C_{p_jp_k}\vec L_{p_k}
  \right)+\rmi C_{\delta_jp_k}\vec L_{p_k}\right\}
\label{eq:01-99}
\end{equation}
and
\begin{equation}
  Z_{jklm}^{(2)} := \int\rmd\bi q\,
  \e^{\rmi\left\langle\bi L_q,\bi q\right\rangle}
  \left(\vec L_{p_j}^\top C_{p_jp_k}\vec L_{p_k}\right)
  \left(\vec L_{p_l}^\top C_{p_lp_m}\vec L_{p_m}\right)\;.
\label{eq:01-100}
\end{equation}
Note that the $Z_{jk}^{(1)}$ are not necessarily symmetric in $(j,k)$ because of the $C_{\delta_jp_k}$ correlation between densities and momenta. In contrast, the $Z_{jklm}^{(2)}$ are symmetric under the permutations $(jklm)\to(lmjk)$, $(jklm)\to (kjlm)$ and $(jklm)\to(jkml)$.

\subsection{Linear momentum correlations}
\label{ssc:4.7}

With the explicit expressions (\ref{eq:A01-21}), (\ref{eq:A01-22}) and (\ref{eq:A01-23}) for the components of $C_{\delta_j\delta_k}$, $C_{\delta_jp_k}$ and $C_{p_jp_k}$, and using that the power spectra for the density and the velocity potential are related by (\ref{eq:A01-6})
\begin{equation}
  P_\delta(\vec k\,) = k^4\,P_\psi(\vec k\,)\;,
\label{eq:01-101}
\end{equation}
we immediately obtain from (\ref{eq:01-99}) the result
\begin{equation}
  Z_{jk}^{(1)} =
  (2\pi)^3\delta_\mathrm{D}\left(\vec L_{q_j}+\vec L_{q_k}\right)
  \mathcal{N}'_{jk}
  P_\delta\left(\vec L_{q_j}\right)A_{jk}^2\left(\vec L_{q_j}\right)\;,
\label{eq:01-102}
\end{equation}
where the abbreviations
\begin{equation}
  A_{jk}^2\left(\vec L_{q_j}\right) := \frac{1}{2}\left(
    1-a_{jk}^2\left(\vec L_{q_j}\right)
  \right)-b_{jk}\left(\vec L_{q_j}\right)
\label{eq:01-103}
\end{equation}
with
\begin{equation}
  a_{jk}^2\left(\vec L_{q_j}\right) :=
  \frac{\left(\vec L_{p_j}\cdot\vec L_{q_j}\right)
        \left(\vec L_{q_j}\cdot\vec L_{p_k}\right)}
  {\vec L_{q_j}^{\,4}} \;,\quad
  b_{jk}\left(\vec L_{q_j}\right) :=
  \frac{\vec L_{q_j}\cdot\vec L_{p_k}}{\vec L_{q_j}^{\,2}}\;,
\label{eq:01-104}
\end{equation}
as well as
\begin{equation}
  \mathcal{N}'_{jk} := \int\rmd\bi q'\,
  \e^{\rmi\left\langle\bi L_q,\bi q'\right\rangle}
\label{eq:01-105}
\end{equation}
were defined. The prime on $\mathcal{N}_{jk}'$ indicates that $\vec q_j$ and $\vec q_k$ are excluded from $\bi q$ here.

\subsection{Quadratic momentum correlations}
\label{ssc:4.8}

For taking quadratic initial momentum-momentum correlations into account, we need to evaluate different terms contributing to $Z_{jklm}^{(2)}$ in (\ref{eq:01-100}). In view of the symmetries of $Z_{jklm}^{(2)}$, we find it convenient to distinguish terms with two equal index pairs, $Z_{jkjk}^{(2)}$, terms with one double index, $Z_{jkkl}^{(2)}$, and terms with four different indices, $Z_{jklm}^{(2)}$.

For two equal index pairs, we find
\begin{equation}
  \label{eq:01-106}
  Z_{jkjk}^{(2)} = (2\pi)^3
  \delta_\mathrm{D}\left(\vec L_{q_j}+\vec L_{q_k}\right)\mathcal{N}'_{jk}
  \int_k
  P_\delta\left(\vec k\,\right)P_\delta\left(\vec k-\vec L_{q_j}\right)
  a_{jk}^2\left(\vec k\,\right)a_{jk}^2\left(\vec k-\vec L_{q_j}\right)\;,
\end{equation}
for one double index,
\begin{equation}
  Z_{jkkl}^{(2)} = (2\pi)^3
  \delta_\mathrm{D}\left(\vec L_{q_j}+\vec L_{q_k}+\vec L_{q_l}\right)
  \mathcal{N}'_{jkl}
  P_\delta\left(\vec L_{q_j}\right)P_\delta\left(\vec L_{q_l}\right)
  a_{jk}^2\left(\vec L_{q_j}\right)a_{kl}^2\left(\vec L_{q_l}\right)\;,
\label{eq:01-107}
\end{equation}
and for four different indices,
\begin{equation}
  \fl
  Z_{jklm}^{(2)} =
  (2\pi)^3\delta_\mathrm{D}\left(\vec L_{q_j}+\vec L_{q_k}\right)
  (2\pi)^3\delta_\mathrm{D}\left(\vec L_{q_l}+\vec L_{q_m}\right)
  \mathcal{N}'_{jklm}
  P_\delta\left(\vec L_{q_j}\right)P_\delta\left(\vec L_{q_l}\right)
  a_{jk}^2\left(\vec L_{q_j}\right)a_{lm}^2\left(\vec L_{q_l}\right)\;.
\label{eq:01-108}
\end{equation}
Given those expressions, the free generating functional is approximated by (\ref{eq:01-95}) and (\ref{eq:01-96}). Further progress can be made once the shift tensor $\bi L$ is specified, for example when the density correlators are to be calculated; see Sect.~VI below.

\section{First-order perturbation theory}
\label{sc:5}

\subsection{One- and two-point correlators with first-order interaction}
\label{ssc:5.1}

We have shown in (\ref{eq:01-58}) that the generating functional including interaction can be created from the free generating functional $Z_0[\bi J, \bi K]$ by means of an interaction operator,
\begin{equation}
  Z[H,\bi J,\bi K] =
  \e^{\rmi\hat S_\mathrm{I}}\e^{\rmi H\cdot\hat\Phi}Z_0[\bi J,\bi K]\;,
\label{eq:01-109}
\end{equation}
with the interaction part of the action given by the operator
\begin{equation}
  \hat S_\mathrm{I} = -\int\rmd 1\left(
    \frac{\delta}{\rmi\delta H_B(-1)}v(1)\frac{\delta}{\rmi\delta H_\rho(1)}
  \right)\;,
\label{eq:01-110}
\end{equation}
defined with slightly more explicit notation in (\ref{eq:01-57}). As we have noted before, this expression for the interaction part of the action contains the two assumptions that the potential is assumed to be translation invariant and acts instantaneously.

Since the functional derivatives with respect to $H$ in (\ref{eq:01-110}) act only on the collective-field operator $\e^{\rmi H\cdot\hat\Phi}$, the effect of the interaction operator can be brought into the form
\begin{equation}
  Z[H,\bi J,\bi K] = \e^{\rmi H\cdot\hat\Phi}
  \e^{\rmi S_\mathrm{I}}Z_0[\bi J,\bi K]
\label{eq:01-111}
\end{equation}
with
\begin{equation}
  \hat S_\mathrm{I} = -\int\rmd 1\,
    \hat\Phi_B(-1)\,v(1)\,\hat\Phi_\rho(1)\;.
\label{eq:01-112}
\end{equation}
The density and response-field operators, $\hat\Phi_\rho$ and $\hat\Phi_B$, in the interaction part $S_\mathrm{I}$ of the action now act directly on the free generating functional and produce correlators introduced in (\ref{eq:01-72}) before. To lowest non-trivial order, the interaction operator is
\begin{equation}
  \e^{\rmi\hat S_\mathrm{I}} \approx 1-
  \rmi\int\rmd 1\hat\Phi_B(-1)\,v(1)\,\hat\Phi_\rho(1)\;.
\label{eq:01-113}
\end{equation}

The corrections to the one- and two-point density correlators in first non-trivial order are then
\begin{equation}
  \delta^{(1)}G_\rho(1) =
  \hat\Phi_\rho(1)\left(-\rmi\hat S_\mathrm{I}Z_0[\bi J,\bi K]\right) =
  -\rmi\int\rmd1'\,v(1')\,G_{B\rho\rho}({-1'}1'1)
\label{eq:01-114}
\end{equation}
and similarly
\begin{equation}
  \delta^{(1)}G_{\rho\rho}(12) =
  -\rmi\int\rmd 1'\,v(1')\,G_{B\rho\rho\rho}({-1'}1'21)\;.
\label{eq:01-115}
\end{equation}
Note that we mark with primes the internal vertices of the interaction, which are integrated over in the interaction operator.

For calculating the first-order approximation of the non-linear density evolution and the non-linear power spectrum, we thus have to work out the three- and four-point correlators $G_{B\rho\rho}({-1'}1'1)$ and $G_{B\rho\rho\rho}({-1'}1'21)$ from the free generating functional.

In view of our later cosmological application, we anticipate that the potential satisfies a Poisson equation of the form
\begin{equation}
  \nabla^2v\left(\vec q, t\right) = g_v(t)\delta
\label{eq:01-116}
\end{equation}
with a function $g_v(t)$ to be specified, where $\delta$ is again the number-density contrast of the particles. Since we are here aiming at the potential caused by a single particle, we can place this particle without loss of generality into the origin of a coordinate system and write its contribution to the density contrast as
\begin{equation}
  \delta = \bar\rho^{-1}\delta_\mathrm{D}\left(\vec q\,\right)-1\;,
\label{eq:01-117}
\end{equation}
with $\bar\rho$ being the mean particle number density. Fourier transforming (\ref{eq:01-116}) then gives
\begin{equation}
  v(1) = -\frac{g_v(t_1)}{k_1^2}\left(\frac{1}{\bar\rho}-\hat 1\right)\;.
\label{eq:01-118}
\end{equation}
The Fourier-transformed unity $\hat 1$ can be neglected later because the zero mode of the potential will not contribute to any correlators. We can thus insert
\begin{equation}
  v(1) = -\frac{g_v(t_1)}{\bar\rho k_1^2}
\label{eq:01-119}
\end{equation}
for the Fourier-transformed, one-particle potential $v$. Notice in particular that this potential scales inversely with the mean particle density $\bar\rho$. This is because, for a fixed mean mass per volume, the particle mass has to decrease in inverse proportion to the particle number $N$ if that number is increased.

\subsection{Shot noise and the relevance of terms}
\label{ssc:5.2}

In our microscopic approach, shot-noise terms appear because the density field is composed of discrete particles. To identify these terms and to clarify their relevance, consider a statistically homogeneous density field
\begin{equation}
  \rho\left(\vec q\,\right) = \sum_{i=1}^N\delta_\mathrm{D}\left(\vec q-\vec q_i\right)
\label{eq:01-120}
\end{equation}
composed of $N$ point particles. In Fourier space, this density field is
\begin{equation}
  \rho\left(\vec k\,\right) = \sum_{i=1}^N\e^{-\rmi\vec k\cdot\vec q_i}\;.
\label{eq:01-121}
\end{equation}
In terms of the density contrast $\delta$, the power spectrum of a continuous density field is
\begin{equation}
  \left\langle
    \rho\left(\vec k\,\right)\rho\left(\vec k^{\,\prime}\right)
  \right\rangle =
  \bar\rho^2\left(
    \hat 1+\left\langle
      \delta\left(\vec k\,\right)\delta\left(\vec k^{\,\prime}\right)
    \right\rangle
  \right) =
  \bar\rho^2\left(
    \hat 1+(2\pi)^3\delta_\mathrm{D}\left(\vec k+\vec k'\right)
    P_\delta\left(k\right)
  \right)
\label{eq:01-122}
\end{equation}
by definition of the density-contrast power spectrum $P_\delta(k)$. If the density fluctuations are uncorrelated,
\begin{equation}
  \left\langle\rho\rho'\right\rangle = \bar\rho^2\,\hat 1\;.
\label{eq:01-123}
\end{equation}

On the other hand, calculating the variance of (\ref{eq:01-121}) results in
\begin{eqnarray}
  \left\langle\rho\rho'\right\rangle &=
  \left\langle\sum_{i,j=1}^N\e^{-\rmi\vec k\cdot\vec q_i-\rmi\vec k^{\,\prime}\cdot\vec q_j}\right\rangle \nonumber\\ &=
  \left(\prod_{k=1}^N\int\frac{\rmd^3q_k}{V}\right)\left(
    \sum_{i=1}^N\e^{-\rmi(\vec k+\vec k^{\,\prime})\cdot\vec q_i}+
    \sum_{i\ne j=1}^N\e^{-\rmi\vec k\cdot\vec q_j-\rmi\vec k^{\,\prime}\cdot\vec q_j}
  \right) \nonumber\\ &=
  \frac{N}{V}(2\pi)^3\delta_\mathrm{D}\left(\vec k+\vec k^{\,\prime}\right)+
  \frac{N(N-1)}{V^2}(2\pi)^3\delta_\mathrm{D}\left(\vec k\,\right)(2\pi)^3\delta_\mathrm{D}\left(\vec k^{\,\prime}\right) \nonumber\\ &=
  \bar\rho(2\pi)^3\delta_\mathrm{D}\left(\vec k+\vec k^{\,\prime}\right)+\bar\rho^2\hat 1^2\;,
\label{eq:01-124}
\end{eqnarray}
abbreviating the Fourier-transformed unity by $\hat 1$ as in (\ref{eq:01-94}). The final step follows by approximating $N(N-1)\approx N^2$. Obviously, only the second term in (\ref{eq:01-124}) corresponds to the result (\ref{eq:01-123}) for the continuous density field, while the first arises only because the density field is composed of discrete particles. Thus, the first term in (\ref{eq:01-124}) is a shot-noise term which arises from summing over pairs of identical particles, as the calculation shows.

More generally, for $m$-point correlators of density fields composed of discrete particles, an analogous calculation shows that terms proportional to all powers of $\bar\rho$ occur, $\bar\rho^s$, with $1\le s\le m$. Only the term proportional to $\bar\rho^m$ is not a shot-noise term. It is the only term arising from summing over combinations of particles which are all different. Terms proportional to powers of $\bar\rho^s$ with $s<m$ are all shot-noise terms in the sense that they arise because of the discrete nature of the density field. In the thermodynamic limit $N\to\infty$, the shot-noise terms can be neglected relative to the dominant term proportional to $\bar\rho^m$.

In the case of gravitational interaction between the microscopic particles, the interaction potential scales with the particle mass. Resolving the density field into an increasing number of particles while keeping the \emph{mass} density constant, the particle mass must be decreased proportional to $N^{-1}$. This repeats the argument made following (\ref{eq:01-119}): The Poisson equation then implies that the gravitational interaction potential must scale inversely with the mean \emph{number} density of particles, i.e.\ like $\bar\rho^{-1}$.

According to (\ref{eq:01-112}), the interaction operator from the interaction part $S_\mathrm{I}$ of the action increases the order of the density $\rho$ and the response field $B$ in the free correlators by one each and multiplies with a potential. As (\ref{eq:01-71}) shows, the response field identifies two particles, as expressed by the Kronecker symbol $\delta_{j_mj_s}$ there. Comparing this with our earlier result on the origin of shot-noise terms, we see that the identification of particles by the response field only selects shot-noise terms from the free density correlators because the only non-shot noise term in the free density correlators arises from combinations of different particles, for which $\delta_{j_mj_s} = 0$.

Specifically, for an $m$-point density correlator in $n$-th order perturbation theory, free correlators of order up to $m+2n$ need to be calculated which are of $(m+n)$-th order in the density and $n$-th order in the response field. In these free correlators, terms proportional to all powers of $\bar\rho$ up to $\bar\rho^{m+2n}$ will occur. Their subsequent multiplication by $v^n$ will reduce the power of $\bar\rho$ by $n$ to $\bar\rho^{m+n}$. Each response field will identify particles pairwise and will thus further reduce the power of the leading term to $\bar\rho^m$, as expected for an $m$-point density correlator.

This shows that only such terms in the free correlators of order $m+2n$ need to be considered which are proportional to $\bar\rho^{m+n}$. Terms proportional to lower powers of $\bar\rho$ will vanish in the limit $N\gg1$, while terms proportional to higher powers of $\bar\rho$ disappear because of the identification of particles by the response fields.

\section{Low-order correlators}
\label{sc:6}

\subsection{Two-point correlator $G_{\rho\rho}(12)$ from linear momentum correlations}
\label{ssc:6.1}

Since the microscopic particles cannot be distinguished, it suffices to select any set of $m$ out of the $N$ particles to evaluate the remaining sum in (\ref{eq:01-78}). These $m$ particles can be labelled with indices from $1$ to $m$ without loss of generality. The generating functional $Z_0[\bi L,0]$ then needs to be calculated for this specific selection of particles, and the resulting terms multiplied by the number of possibilities for the particular subset of $m$ particles selected from the canonical ensemble of $N$ particles.

If we wish to calculate density correlators taking momentum correlations into account to first or second order, we need to evaluate the expressions $Z_{jk}^{(1)}$ from (\ref{eq:01-102}) and $Z_{jklm}^{(2)}$ given in (\ref{eq:01-102}), (\ref{eq:01-107}) and (\ref{eq:01-108}).

For a two-point correlator, $m = 2$, we can choose $j_1,j_2 \in \{1,2\}$. Since the particles have to be different for the correlation terms in $Z_{jk}^{(1)}$ and $Z_{jklm}^{(2)}$ not to vanish, we set $(j_1,j_2) = (1,2)$. Then,
\begin{equation}
  \vec L_{q_1} = -\vec k_1\;,\quad\vec L_{q_2} = -\vec k_2\;,
\label{eq:01-125}
\end{equation}
accordingly
\begin{equation}
  \vec L_{p_1} = -\vec T_1\;,\quad
  \vec L_{q_2} = -\vec T_2\;,
\label{eq:01-126}
\end{equation}
and no other components of $\bi L_{q,p}$ appear. Then, the remaining integrals over all positions except $\vec q_1$ and $\vec q_2$ simply give
\begin{equation}
  \mathcal{N}_{12}' = V^{N-2}\;.
\label{eq:01-127}
\end{equation}
If we set $t_1 = t_2$, i.e.\ if the correlator is taken synchronously, we further have
\begin{equation}
  A_{12}^2\left(\vec k_1\right) = \frac{1}{2}\left(
    1+g_{qp}^2(t_1,0)
  \right)+g_{qp}(t_1,0) =
  \frac{1}{2}\left(1+g_{qp}(t_1,0)\right)^2\;,
\label{eq:01-128}
\end{equation}
taking into account that the remaining delta distribution ensures $\vec k_1 = -\vec k_2$. Therefore,
\begin{equation}
  Z_{12}^{(1)} = \frac{V^{N-2}}{2}
  (2\pi)^3\delta_\mathrm{D}\left(\vec k_1+\vec k_2\right)
  \left(1+g_{qp}(t_1,0)\right)^2P_\delta\left(\vec k_1\right)\;.
\label{eq:01-129}
\end{equation}
This expression is symmetric under the permutation $(1,2)\to(2,1)$. Since the index pair $(1,2)$ can be selected in $N(N-1)\approx N^2$ ways from the $N$ particles, we immediately find
\begin{equation}
  G_{\rho\rho}^{(1)}(12) = \bar\rho^2
  (2\pi)^3\delta_\mathrm{D}\left(\vec k_1+\vec k_2\right)
  \left(1+g_{qp}(t_1,0)\right)^2P_\delta\left(\vec k_1\right)\;.
\label{eq:01-130}
\end{equation}

\subsection{Two-point correlator $G_{\rho\rho}(12)$ from quadratic momentum correlations}
\label{ssc:6.2}

Proceeding to the contribution of quadratic momentum correlations to the two-point correlator, we see immediately that only terms of the form
\begin{equation}
  Z_{jkjk}^{(2)} = (2\pi)^3
  \delta_\mathrm{D}\left(\vec L_{q_j}+\vec L_{q_k}\right)\mathcal{N}'_{jk}
  \int_k
  P_\delta\left(\vec k\,\right)P_\delta\left(\vec k-\vec L_{q_j}\right)
  a_{jk}^2\left(\vec k\,\right)a_{jk}^2\left(\vec k-\vec L_{q_j}\right)
\label{eq:01-131}
\end{equation}
derived in (\ref{eq:01-106}) can contribute because terms with three or four different particle indices must vanish for a two-point correlator. Setting again $(j_1,j_2)=(1,2)$, and evaluating the factors $a_{jk}^2 = a_{12}^2$ with the appropriate momentum shift vectors (\ref{eq:01-126}), we immediately arrive at
\begin{eqnarray}
  Z_{1212}^{(2)} &= V^{N-2}
  (2\pi)^3\delta_\mathrm{D}\left(\vec k_1+\vec k_2\right)
  g_{qp}^4(t_1,0) \nonumber\\ &\cdot
  \int_k
  P_\delta\left(\vec k\,\right)P_\delta\left(\vec k-\vec k_1\right)
  \left(\frac{\vec k_1\cdot\vec k}{k^2}\right)^2
  \left(
    \frac{\vec k_1\cdot(\vec k-\vec k_1)}{(\vec k-\vec k_1)^2}
  \right)^2\;.
\label{eq:01-132}
\end{eqnarray}

As discussed below (\ref{eq:01-100}), the terms $Z_{jklm}^{(2)}$ are symmetric in the first and second index pairs and under exchanges of the two index pairs, and the indices in the first and in the second index pairs must be different. Under these requirements, the term (\ref{eq:01-132}) appears $4$ times in the sum over particle indices: terms with the index combinations $(1212)$, $(2112)$, $(1221)$ and $(2121)$ are all equivalent, and others do not appear. Furthermore, we have to multiply with the number $N(N-1) \approx N^2$ of ways for selecting a pair from the $N$ particles. Thus, by (\ref{eq:01-98}), we arrive at the contribution
\begin{eqnarray}
  G_{\rho\rho}^{(2)}(12) &= \frac{\bar\rho^2}{2}
  (2\pi)^3\delta_\mathrm{D}\left(\vec k_1+\vec k_2\right)
  g_{qp}^4(t_1,0) \nonumber\\ &\cdot
  \int_k
  P_\delta\left(\vec k\,\right)P_\delta\left(\vec k-\vec k_1\right)
  \left(\frac{\vec k_1\cdot\vec k}{k^2}\right)^2
  \left(
    \frac{\vec k_1\cdot(\vec k-\vec k_1)}{(\vec k-\vec k_1)^2}
  \right)^2
\label{eq:01-133}
\end{eqnarray}
of quadratic momentum correlations to the two-point density correlator. The damping term is set to unity here as discussed in Subsect.~\ref{ssc:4.4} before.

\subsection{One- and two-point response-field correlators}
\label{ssc:6.3}

The effect of a single, one-particle response-field operator on the free generating functional was shown in (\ref{eq:01-53}). That expression, valid for a single response-field operator applied to the free generating functional, is easily generalised. Suppose we apply $m$ operators in total, of which $n$ are density and $m-n$ are response-field operators. Since each response-field operator contains a density operator to be executed first, we will have to apply $m$ density operators in total. The result will have to be multiplied by $m-n$ response-field factors. Thus, we have
\begin{eqnarray}
  &\underbrace{\hat\Phi_B(m)\ldots\hat\Phi_B(m-n)}_{m-n\;\mathrm{terms}}
  \underbrace{\hat\Phi_\rho(n)\ldots\hat\Phi_\rho(1)}_{n\;\mathrm{terms}}\,
  \left.Z_0[\bi J, \bi K]\right\vert_{\bi J=0=\bi K} \nonumber\\
  &= \hat b(m)\ldots\hat b(m-n)\hat\Phi_\rho(m)\ldots\hat\Phi_\rho(1)\,
  \left.Z_0[\bi J,\bi K]\right\vert_{\bi J=0=\bi K} \nonumber\\
  &= \hat b(m)\ldots\hat b(m-n)\,
  \left.Z_0[\bi L,\bi K]\right\vert_{\bi K=0}\;.
\label{eq:01-134}
\end{eqnarray}

Decomposing the density and the response-field operators into their single-particle contributions, we first obtain the shift tensor $\bi L$ from (\ref{eq:01-70}). Then, any single-particle response-field operator $\hat b_{j_l}(l)$ returns the factor
\begin{equation}
  b_{j_l}(l) = \rmi\sum_{s=1}^mg_{qp}(t_s,t_l)\,\vec k_s\cdot\vec k_l\,
  \delta_{j_lj_s}\;.
\label{eq:01-135}
\end{equation}

Applying a single response-field operator to the generating functional, we obtain the mean response field. In the general approach outlined above, we set $m = 1$ and $n = 0$. Then, from (\ref{eq:01-135}), we have
\begin{equation}
  b_{j_1}(1) = \rmi g_{qp}(t_1, t_1)\,k_1^2 = 0
\label{eq:01-136}
\end{equation}
if the propagator $g_{qp}(t,t')$ vanishes for $t = t'$, as it usually will. Then, the mean response field vanishes identically.

For $m = 2$, we have the density-response correlator
\begin{equation}
  G_{\rho_{j_2}B_{j_1}}(12) =
  \rmi\vec k_1\cdot\left(
    g_{qp}(t_1,t_1)\vec k_1\delta_{j_1j_1}+
    g_{qp}(t_1,t_2)\vec k_2\delta_{j_1j_2}
  \right)G_{\rho_{j_1}\rho_{j_2}}(12)
\label{eq:01-137}
\end{equation}
according to (\ref{eq:01-135}). Since $g_{qp}(t_1,t_1) = 0$, the first term in parentheses vanishes. The second term contributes only if $j_1 = j_2$ because of the Kronecker symbol, but then the correlation terms in (\ref{eq:01-96}) cannot contribute. Since
\begin{equation}
  \mathcal{N} = (2\pi)^3\delta_\mathrm{D}\left(\vec k_1+\vec k_2\right)V^{N-1}
\label{eq:01-138}
\end{equation}
in this case and
\begin{equation}
  Q_D = \frac{\sigma_1^2}{3}\left(
    g_{qp}(t_1,0)\vec k_1+g_{qp}(t_2,0)\vec k_2
  \right)^2 =
  \frac{\sigma_1^2}{3}\left(g_{qp}(t_1,0)-g_{qp}(t_2,0)\right)^2k_1^2\;,
\label{eq:01-139}
\end{equation}
taking $\vec k_2=-\vec k_1$ into account, we find
\begin{equation}
  \fl
  G_{\rho B}(12) = -\rmi\bar\rho(2\pi)^3
  \delta_\mathrm{D}\left(\vec k_1+\vec k_2\right)\,g_{qp}(t_1,t_2)k_1^2
  \exp\left(
    -\frac{\sigma_1^2k_1^2}{6}
    \left(g_{qp}(t_1,0)-g_{qp}(t_2, 0)\right)^2
  \right)
\label{eq:01-140}
\end{equation}
after summing over all $1\le j_1\le N$. Changing the order of $B$ and $\rho$ in (\ref{eq:01-137}) only changes the ordering of the times $t_1$ and $t_2$, thus leading to
\begin{equation}
  \fl
  G_{B\rho}(12) = -\rmi\bar\rho(2\pi)^3
  \delta_\mathrm{D}\left(\vec k_1+\vec k_2\right)\,g_{qp}(t_2,t_1)k_1^2
  \exp\left(
    -\frac{\sigma_1^2k_1^2}{6}
    \left(g_{qp}(t_1,0)-g_{qp}(t_2, 0)\right)^2
  \right)\;.
\label{eq:01-141}
\end{equation}
The cross-spectra $G_{\rho B}(12)$ and $G_{B\rho}(12)$ will obviously vanish if $t_1 = t_2$. Applying both $\hat b_{j_1}(1)$ and $\hat b_{j_2}(2)$ will return a product of propagators with time orderings $(t_1,t_2)$ and $(t_2,t_1)$, which must vanish for causality, hence
\begin{equation}
  G_{BB}(12) = 0\;.
\label{eq:01-142}
\end{equation}

\subsection{Three-point correlator $G_{B\rho\rho}({-1'}1'1)$ from linear and quadratic momentum correlations}
\label{ssc:6.4}

We shall now proceed to work out the three- and four-point correlators $G_{B\rho\rho}({-1'}1'1)$ and $G_{B\rho\rho\rho}({-1'}1'21)$ we require. For all calculations to be carried out below, it is important that the response field identifies two particles, which is mathematically expressed by the Kronecker delta in (\ref{eq:01-135}). Effectively, therefore, $m$-point correlators of the form $G_{B\rho\ldots\rho}$ identify two particles. Accordingly, in the three- and four-point correlators that we are about to calculate, only two and three particles are free, respectively. Since these particles are indistinguishable, we can enumerate them with indices $(j_1,j_2) = (1,2)$ and $(j_1,j_2,j_3)=(1,2,3)$ and multiply the results with the number of ways to choose particle pairs and particle triples from an ensemble of $N$ particles.

We begin with the correlators derived from the generating functional $Z_0^{(1)}[\bi L,0]$ from (\ref{eq:01-96}), which contains momentum correlations to linear order only. For $m = 3$, the one-particle response-field factor in (\ref{eq:01-135}) reduces to the single term
\begin{equation}
  b_{j_2'}(2') = -\rmi\,g_{qp}(t_1,t_1')\,\vec k_1'\cdot\vec k_1\,
  \delta_{j_1j_2'}
\label{eq:01-143}
\end{equation}
because $t_1' = t_2'$ and therefore $g_{qp}(t_1',t_2') = g_{qp}(t_2',t_2') = 0$. Moreover, we have replaced $\vec k_2'$ by $-\vec k_1'$, expressing the translation invariance of the potential $v$. Since the Kronecker symbol in the response-field factor identifies the particles $j_1$ and $j_2'$, only two particle indices are free, which we set without loss of generality to $(j_1,j_1')=(1,2)$. The shift vectors $\vec L_{q_j}$ are then
\begin{equation}
  \vec L_{q_j} = -\left(\vec k_1-\vec k_1'\right)\delta_{j1}-\vec k_1'\delta_{j2}\;.
\label{eq:01-144}
\end{equation}
For the two-particle term (\ref{eq:01-102}), we can label the two particles by $(j,k)=(1,2)$ and thus write
\begin{equation}
  \vec L_{q_1} = -\left(\vec k_1-\vec k_1'\right)\;,\quad
  \vec L_{q_2} = -\vec k_1'\;.
\label{eq:01-145}
\end{equation}
We can stop here: The delta distribution in the two-particle term in (\ref{eq:01-102}) shrinks to
\begin{equation}
  \delta_\mathrm{D}\left(\vec L_{q_1}+\vec L_{q_2}\right) = \delta_\mathrm{D}\left(\vec k_1\right)
\label{eq:01-146}
\end{equation}
and ensures this way that $\vec k_1 = 0$, which sets the response-field factor (\ref{eq:01-143}) to zero. We can thus conclude that $G_{B\rho\rho}({-1'}1'1)$ cannot contribute at all to the one-point correlator, hence
\begin{equation}
  \delta^{(1)}G_\rho(1) = 0
\label{eq:01-147}
\end{equation}
to first order in the interaction and to linear order in the momentum correlations: To this order, the interaction does not change the mean density.

For the two-particle term (\ref{eq:01-106}) contributing to the quadratic momentum correlation, we can also set $(j,k)=(1,2)$ and arrive at the same conclusion: The delta distribution ensures $\vec k_1 = 0$ and thus sets the response to zero. The three-particle term (\ref{eq:01-107}) cannot contribute because $\vec L_{q_3} = 0$ according to (\ref{eq:01-144}), which implies $a_{23}^2 = 0$.

Of course, this is not surprising: No internal interaction between identical particles can change the mean density in a statistically homogeneous, canonical ensemble. It is merely reassuring to see why the individual contributions disappear formally.

\subsection{Three-point correlator $G_{\rho\rho\rho}(123)$ from linear and quadratic momentum correlations}
\label{ssc:6.5}

The contribution to the three-point correlator $G_{\rho\rho\rho}(123)$ due to linear momentum correlations can be easily read off (\ref{eq:01-96}). Ignoring the damping term and focussing on the correlated contribution to the free generating functional $Z_0^{(1)}[\bi L,0]$ in (\ref{eq:01-96}), we first notice that again neither $\vec L_{q_1}$ nor $\vec L_{q_2}$ must vanish
because otherwise individual wave vectors would be set to zero, causing the power spectrum to disappear. Therefore, at least one each of the particle indices $(j_1,j_2,j_3)$ must be set to $1$ and $2$, while the third particle index available for the three-point correlator remains free.

If we set this third index to $1$ or $2$ as well, the multiplicity of the resulting term is $\propto N^2$, which is lower by a factor of $N$ than the multiplicity $\propto N^3$ required for the three-point correlator. This term is thus negligible. Only terms with the third index set to $>2$ will remain. Adopting $(j_1,j_2,j_3) = (1,2,3)$ implies
\begin{equation}
  \vec L_{q_1} = -\vec k_1\;,\quad
  \vec L_{q_2} = -\vec k_2\;,\quad
  \mathcal{N}_{12}' = (2\pi)^3\delta_\mathrm{D}\left(\vec k_3\right)V^{N-3}\;.
\label{eq:01-148}
\end{equation}
Moreover, for a synchronous three-point correlator, $t_1 = t_2 = t_3$. Thus,
\begin{equation}
  A_{12}^2 = \frac{1}{2}\left(
    1-g_{qp}(t_1,0)^2\frac{\vec k_1\cdot\vec k_2}{k_1^2}
  \right)-g_{qp}(t_1,0)\frac{\vec k_1\cdot\vec k_2}{k_1^2} =
  \frac{1}{2}\left(1+g_{qp}(t_1,0)\right)^2\;,
\label{eq:01-149}
\end{equation}
where the latter step follows because one of the remaining delta distributions ensures $\vec k_1=-\vec k_2$. Combining results, we find
\begin{equation}
  Z_{12}^{(1)} = \frac{V^{N-3}}{2}
  (2\pi)^6\delta_\mathrm{D}\left(\vec k_1+\vec k_2\right)
  \delta_\mathrm{D}\left(\vec k_3\right)
  \left(1+g_{qp}(t_1,0)\right)^2\,P_\delta\left(\vec k_1\right)\;.
\label{eq:01-150}
\end{equation}
The index combination $(j_1,j_2,j_3) = (2,1,3)$ adds the same expression. Taking the remaining cyclic index permutations into account leads to the contribution
\begin{equation}
  G_{\rho\rho\rho}^{(1)}(123) = \bar\rho^3
  (2\pi)^6\left(1+\tau_1\right)^2 
  \left\{
    \delta_\mathrm{D}\left(\vec k_1+\vec k_2\right)
    \delta_\mathrm{D}\left(\vec k_3\right)
    P_\delta\left(\vec k_1\right) + \mbox{cyc.}
  \right\}
\label{eq:01-151}
\end{equation}
to the three-point correlator from linear momentum correlations.

The terms of second order in the momentum correlation can be read off (\ref{eq:01-106}) and (\ref{eq:01-107}). The two-particle term in $Z_0^{(2)}[\bi L,0]$ gives
\begin{eqnarray}
  Z_{1212}^{(2)} &= V^{N-3}(2\pi)^6
  \delta_\mathrm{D}\left(\vec k_1+\vec k_2\right)
  \delta_\mathrm{D}\left(\vec k_3\right)g_{qp}^4(t_1,0) \nonumber\\ &\cdot
  \int_kP_\delta\left(\vec k\right)
  P_\delta\left(\vec k-\vec k_1\right)
  \left(\frac{\vec k_1\cdot\vec k}{k^2}\right)
  \left(\frac{\vec k_1\cdot(\vec k-\vec k_1)}
             {(\vec k-\vec k_1)^2}\right)\;.
\label{eq:01-152}
\end{eqnarray}
Since there are again four equivalent index configurations for this term, and since the three index combinations $(1,2)$, $(1,3)$ and $(2,3)$ are possible for the three-point correlator, we arrive at the contribution
\begin{equation}
  G_{\rho\rho\rho}^{(2A)}(123) =
  \bar\rho(2\pi)^3\left\{
    \delta_\mathrm{D}\left(\vec k_3\right)G_{\rho\rho}^{(2)}(12) +
    \mbox{cyc.}
  \right\}
\label{eq:01-153}
\end{equation}
of quadratic momentum correlations to the three-point correlator, with $G_{\rho\rho}^{(2)}(12)$ taken from (\ref{eq:01-133}).

Finally, the three-particle term $Z_{jkkl}^{(2)}$ from (\ref{eq:01-107}) gives
\begin{eqnarray}
  Z_{1223}^{(2)} &=
  V^{N-3}(2\pi)^3\delta_\mathrm{D}\left(\vec k_1+\vec k_2+\vec k_3\right)
  g^4_{qp}(t_1,0)
  P_\delta\left(\vec k_1\right)P_\delta\left(\vec k_3\right)
  \frac{\vec k_1\cdot\vec k_2}{k_1^2}
  \frac{\vec k_2\cdot\vec k_3}{k_3^2}\;.
\label{eq:01-154}
\end{eqnarray}
Due to the symmetries of the $Z_{jklm}^{(2)}$ terms, there are $8$ equivalent index combinations. This term thus contributes
\begin{eqnarray}
  G_{\rho\rho\rho}^{(2B)}(123) &= \bar\rho^3(2\pi)^3
  \delta_\mathrm{D}\left(\vec k_1+\vec k_2+\vec k_3\right)g^4_{qp}(t_1,0)
  \nonumber\\ &\cdot \left\{
    P_\delta\left(\vec k_1\right)P_\delta\left(\vec k_3\right)
    \frac{\vec k_1\cdot\vec k_2}{k_1^2}
    \frac{\vec k_2\cdot\vec k_3}{k_3^2} + \mbox{cyc.}
  \right\}
\label{eq:01-155}
\end{eqnarray}
to the three-point correlator. Since there are no interactions included at this level, we set the damping factor to unity.

\subsection{Four-point correlator $G_{B\rho\rho\rho}({-1'}1'21)$ from linear momentum correlations}
\label{ssc:6.6}

Turning to the effect of first-order interactions on the density power spectrum, we need to work out the four-point correlator $G_{B\rho\rho\rho}({-1'}1'21)$. The response-field factor is
\begin{eqnarray}
  b_{j_2'}(2') &=
  -\rmi g_{qp}(t_1,t_1')\left(
    \vec k_1\cdot\vec k_1^{\,\prime}\,\delta_{j_1j_2'}+
    \vec k_2\cdot\vec k_1^{\,\prime}\,\delta_{j_2j_2'}
  \right)\;,
\label{eq:01-156}
\end{eqnarray}
setting $\vec k_2'=-\vec k_1'$ again. Other terms do not appear here because $g_{qp}(t_1',t_2') = 0 = g_{qp}(t_2',t_2')$. We shall further consider synchronous correlations only and thus set $t_1 = t_2$. Of the two terms remaining in (\ref{eq:01-156}), we now focus on the first, in which the Kronecker symbol ensures that $j_1 = j_2'$. The second term will then be obtained from the result by interchanging the indices $j_1$ and $j_2$ or, equivalently, the wave vectors $\vec k_1$ and $\vec k_2$.

Due to the coupling of two particles, three particles remain free, for which we choose the indices $(j_1,j_2,j_1')=(1,2,3)$ without loss of generality. The shift vectors are then
\begin{equation}
  \vec L_{q_j} = -\left(\vec k_1-\vec k_1'\right)\delta_{j1}-
  \vec k_2\delta_{j2}-\vec k_1'\delta_{j3}\;.
\label{eq:01-157}
\end{equation}

The three particles need to be placed on three different positions to achieve the largest possible multiplicity. We choose three positions labelled by $(j,k,l)=(2,3,1)$, obtain the shift vectors
\begin{equation}
  \vec L_{q_j} = -\vec k_2\;,\quad
  \vec L_{q_k} = -\vec k_1'\;,\quad
  \vec L_{q_l} = -\left(\vec k_1-\vec k_1'\right)
\label{eq:01-158}
\end{equation}
from (\ref{eq:01-157}) and
\begin{equation}
  Z_{23}^{(1)} =
  (2\pi)^3\delta_\mathrm{D}\left(\vec k_2+\vec k_1'\right)
  (2\pi)^3\delta_\mathrm{D}\left(\vec k_1-\vec k_1'\right)
  P_\delta\left(\vec k_2\right)A_{jk}^2\left(\vec k_2\right)
\label{eq:01-159}
\end{equation}
from (\ref{eq:01-102}). The second delta distribution arises from the factor $\mathcal{N}_{jk}'$ in (\ref{eq:01-105}). Since it ensures $\vec k_1'=\vec k_1$, it allows us to write (\ref{eq:01-159}) as
\begin{equation}
  Z_{23}^{(1)} =
  (2\pi)^3\delta_\mathrm{D}\left(\vec k_1+\vec k_2\right)
  (2\pi)^3\delta_\mathrm{D}\left(\vec k_1-\vec k_1'\right)
  P_\delta\left(\vec k_1\right)A_{jk}^2\left(\vec k_1\right)\;,
\label{eq:01-160}
\end{equation}
where $A^2_{jk}(\vec k_1)$ simplifies to
\begin{equation}
  A_{jk}^2 = \frac{1}{2}\left(
    1+g_{qp}(t_1,0)g_{qp}(t_1',0)
  \right)+g_{qp}(t_1',0)
\label{eq:01-161}
\end{equation}
because $\vec k_1 = \vec k_1' = -\vec k_2$ due to the delta distributions. For permutations of $(j,k,l)$ with $l\ne1$, the factor $\mathcal{N}_{jk}'$ results in a delta distribution setting one individual wave vector to zero, which causes the result to vanish. The only other permutation leading to a non-vanishing result is thus $(j,k,l)=(3,2,1)$, for which
\begin{equation}
  A_{jk}^2 = \frac{1}{2}\left(
    1+g_{qp}(t_1,0)g_{qp}(t_1',0)
  \right)+g_{qp}(t_1,0)\;.
\label{eq:01-162}
\end{equation}

After collecting results, the summation over particle indices multiplies the result by $N(N-1)(N-2)\approx N^3$, and the relevant three-particle contribution to the four-point density correlator turns out to be
\begin{eqnarray}
  G^{(1)}_{\rho\rho\rho\rho}({-1'}1'21) &=
  \e^{-Q_D/2}\bar\rho^3(2\pi)^6\delta_\mathrm{D}\left(\vec k_1+\vec k_2\right)
  \delta_\mathrm{D}\left(\vec k_1-\vec k_1'\right) \nonumber\\ &\cdot
  \left(1+g_{qp}(t_1',0)\right)\left(1+g_{qp}(t_1,0)\right)
  P_\delta\left(\vec k_1\,\right)\;.
\label{eq:01-163}
\end{eqnarray}

Recall that this result was obtained assuming $j_1 = j_2'$. It is quite straightforward to see that the contribution for $j_2 = j_2'$ is identical, multiplying the correlator by two. Thus, the four-point correlator required for the first-order perturbation theory according to (\ref{eq:01-115}) is
\begin{eqnarray}
  G_{B\rho\rho\rho}^{(1)}({-1'}1'21) &=
  -2\rmi\,\e^{-Q_D/2}\,g_{qp}(t_1,t_1')
  \left(1+g_{qp}(t_1',0)\right)\left(1+g_{qp}(t_1,0)\right)
  \nonumber\\ &\cdot \bar\rho^3
  (2\pi)^6\delta_\mathrm{D}\left(\vec k_1+\vec k_2\right)
  \delta_\mathrm{D}\left(\vec k_1-\vec k_1'\right)
  k_1^2\,P_\delta\left(\vec k_1\,\right)\;.
\label{eq:01-164}
\end{eqnarray}
With
\begin{equation}
  \bar{\bi L}_p^2 = \sum_{r,s=1}^m\vec T_r\cdot\vec T_s\,\delta_{j_rj_s}\;,
\label{eq:01-165}
\end{equation}
the damping term turns out to be
\begin{equation}
  Q_D = \frac{2\sigma_1^2}{3}\left(
    T_1^2-\vec T_1\cdot\vec T_1'+T_1'^2
  \right)\;.
\label{eq:01-166}
\end{equation}
According to (\ref{eq:01-115}), this implies the contribution
\begin{eqnarray}
\label{eq:01-167}
  \delta^{(1)}G^{(1)}_{\rho\rho}(12) &=
  -2\bar\rho^3(2\pi)^3\delta_\mathrm{D}\left(\vec k_1+\vec k_2\right)
  k_1^2\,P_\delta\left(\vec k_1\,\right) \\ &\cdot
  \int_0^{t_1}\rmd t_1'\,v\left(t_1',\vec k_1\right)
  \e^{-Q_D/2}\,g_{qp}(t_1,t_1')
  \left(1+g_{qp}(t_1',0)\right)\left(1+g_{qp}(t_1,0)\right)
  \nonumber
\end{eqnarray}
to the non-linear power spectrum, where the potential $\hat v(\vec k_1)$ was included in the time integral because its amplitude may depend on time, and the damping term $\e^{-Q_D/2}$ was included there because it does depend on time according to (\ref{eq:01-166}).

\subsection{Four-point correlator $G_{B\rho\rho\rho}({-1'}1'21)$ from quadratic momentum correlations}
\label{ssc:6.7}

We now turn to evaluating the contributions to the density power spectrum from quadratic initial momentum correlations, which are expressed by the free generating functional $Z_0^{(2)}[\bi L,0]$ from (\ref{eq:01-96}). Since the response-field prefactor in (\ref{eq:01-156}) identifies particle pairs, only three particle indices are free, which immediately implies that no four-particle terms can contribute. The two- and three-particle terms from (\ref{eq:01-106}) and (\ref{eq:01-107}) are thus the only ones to consider. Again, we label the particles by $(j_1,j_2,j_1')=(1,2,3)$ without loss of generality.

Regarding the three-particle term $Z_{jkkl}^{(2)}$, the position-index combination $(j,k,l) = (1,2,3)$ leads to
\begin{equation}
  \fl
  Z_{1223}^{(2A)} = \frac{\vec T_1\cdot\vec T_1'}{k_1'^2}
  \frac
   {\vec T_1\cdot(\vec k_1-\vec k_1')
    (\vec T_1-\vec T_1')\cdot(\vec k_1-\vec k_1')}
   {(\vec k_1-\vec k_1')^4}
  (2\pi)^3\delta_\mathrm{D}(\vec k_1+\vec k_2)
  P_\delta\left(\vec k_1-\vec k_1'\right)P_\delta\left(\vec k_1'\right)\;,
\label{eq:01-168}
\end{equation}
the combination $(j,k,l) = (2,3,1)$ gives
\begin{equation}
  \fl
  Z_{1332}^{(2B)} = -\frac{\vec T_1\cdot\vec T_1'}{k_1^2}
  \frac
   {\vec T_1'\cdot(\vec k_1-\vec k_1')
    (\vec T_1-\vec T_1')\cdot(\vec k_1-\vec k_1')}
   {(\vec k_1-\vec k_1')^4}
  (2\pi)^3\delta_\mathrm{D}(\vec k_1+\vec k_2)
  P_\delta\left(\vec k_1\right)P_\delta\left(\vec k_1-\vec k_1'\right)\;,
\label{eq:01-169}
\end{equation}
and the combination $(j,k,l)=(3,1,2)$ produces
\begin{eqnarray}
  Z_{3112}^{(2C)} &= -\frac{\vec T_1\cdot(\vec T_1-\vec T_1')}{k_1^2}
  \frac{\vec T_1'\cdot(\vec T_1-\vec T_1')}{k_1'^2} %\nonumber\\ &\cdot
  (2\pi)^3\delta_\mathrm{D}(\vec k_1+\vec k_2)
  P_\delta\left(\vec k_1\right)P_\delta\left(\vec k_1'\right)\;.
\label{eq:01-170}
\end{eqnarray}

Finally, for the two-particle term in (\ref{eq:01-106}) to contribute, the factor $\mathcal{N}_{jk}'$ returns a delta distribution for an individual wave number except for the particle-index combinations $(j,k,l)=(2,3,1)$ or $(3,2,1)$. For these,
\begin{equation}
  \fl
  Z_{2323}^{(2D)} =
  \delta_\mathrm{D}\left(\vec k_1-\vec k_1'\right)\,
  \int_k
  P_\delta\left(\vec k_1-\vec k\right)P_\delta\left(\vec k\right)
  \frac{\vec T_1\cdot\vec k\,\vec T_1'\cdot\vec k}{k^4}
  \frac
   {\vec T_1\cdot(\vec k_1-\vec k)\vec T_1'\cdot(\vec k_1-\vec k)}
   {(\vec k_1-\vec k)^4}\;.
\label{eq:01-171}
\end{equation}
For all terms in (\ref{eq:01-168}), (\ref{eq:01-169}), (\ref{eq:01-170}) and (\ref{eq:01-171}), the damping term agrees with (\ref{eq:01-166}).

The expressions (\ref{eq:01-168}), (\ref{eq:01-169}) and (\ref{eq:01-170}) each have the multiplicity $2^3=8$ due to the symmetry of the three-point term (\ref{eq:01-107}), while the expression (\ref{eq:01-171}) has the multiplicity $2^2=4$. Summing over all particle indices further multiplies the results by $N(N-1)(N-2) \approx N^3$. Taking the respective factors into account, we arrive at the relevant contribution
\begin{eqnarray}
  G_{\rho\rho\rho\rho}^{(2)}({-1'}1'21) = \bar\rho^3\e^{-Q_D/2}
  \left(
    Z_{1223}^{(2A)}+Z_{2331}^{(2B)}+Z_{3112}^{(2C)}+
    \frac{Z_{2323}^{(2D)}}{2}
  \right)
\label{eq:01-172}
\end{eqnarray}
to the four-point density correlator.

The contribution $G^{(2)}_{B\rho\rho\rho}({-1'}1'21)$ of these terms to the correlator $G_{B\rho\rho\rho}({-1'}1'21)$ follows again by multiplying with the response-field factor (\ref{eq:01-156}), taking into account that both terms lead to same result. Thus,
\begin{equation}
  G^{(2)}_{B\rho\rho\rho}({-1'}1'21) = -2\rmi g_{qp}(t_1,t_1')\,
  \vec k_1\cdot\vec k_1'\,G^{(2)}_{\rho\rho\rho\rho}({-1'}1'21)\;.
\label{eq:01-173}
\end{equation}
Inserting this into (\ref{eq:01-115}), we find
\begin{equation}
  \delta^{(1)}G_{\rho\rho}^{(2)}(12) =
  -2\int_0^{t_1}\rmd 1'
  g_{qp}(t_1,t_1')\,
  \vec k_1\cdot\vec k_1'\,v(1')G^{(2)}_{\rho\rho\rho\rho}({-1'}1'21)\;.
\label{eq:01-174}
\end{equation}

\section{Cosmological power spectra}
\label{sc:7}

\subsection{Power-spectra contributions from the free generating functional}
\label{ssc:7.1}

All results obtained so far for the generating functional, for the initially correlated phase-space distribution and the low-order density- and response-field correlators are generally valid for systems of classical particles. The free phase-space trajectories of these particles are characterised by a known retarded Green's function and they interact with a two-particle potential $v$.

In this Section, we shall specialise these results to classical point particles in cosmology. The essential difference to common classical particle systems is that space is expanding with time. The physical distance $\vec r$ between any two particles thus grows with time in proportion to a scale factor $a(t)$. The spatial coordinates $\vec q$ are taken to be comoving coordinates, defined by $\vec r = a(t)\vec q$.

The Green's function for particles moving freely in such an expanding space has been derived in \cite{2015PhRvD..91h3524B}. Specifically, the free propagator $g_{qp}(\tau,\tau')$ has been shown to be
\begin{equation}
  g_{qp}(\tau,\tau') = \int_{\tau'}^\tau\frac{\rmd\bar\tau}{g(\bar\tau)}\;,
\label{eq:01-175}
\end{equation}
where $\tau = D_+(a)-1$ was introduced as a time coordinate more convenient than the cosmological time $t$ or the cosmological scale factor $a$. The function $D_+(a)$ is the linear growth factor, describing the increase in density-fluctuation amplitudes as long as they remain linear. The growth factor is assumed to be normalised to unity at the initial time such that $\tau = 0$ initially. Moreover, $g(\tau)$ is defined by
\begin{equation}
  g(\tau) := a^2D_+(a)f(a)\frac{H(a)}{H_\mathrm{i}}\quad\mbox{with}\quad
  f(a) := \frac{\rmd\ln D_+}{\rmd\ln a}\;,
\label{eq:01-176}
\end{equation}
including the Hubble function $H(a)$ and the Hubble constant $H_\mathrm{i}$ at the initial time $\tau = 0$. In (\ref{eq:01-176}), the scale factor $a$ is supposed to be normalised to unity at the initial time, hence $g\to1$ for $\tau\to0$.

As a consequence of the expanding space, the propagator $g_{qp}(\tau,\tau')$ remains finite even for $\tau\to\infty$. For an Einstein-de Sitter universe, $g_{qp}(\tau,0) < 2$. Clearly, therefore, inserting the free Hamiltonian propagator (\ref{eq:01-175}) into the free linear power-spectrum term (\ref{eq:01-130}) cannot reproduce the result well-known from ordinary cosmological perturbation theory that the matter power spectrum evolves linearly as $P_\delta(k) \propto D_+^2(a)$.

We can, however, achieve this behaviour by mimicking the Zel'dovich approximation, which implies extrapolating the first-order solution of Lagrangian Perturbation Theory (LPT) beyond the linear regime. LPT describes the motion of fluid elements in terms of a displacement field $b(t)\vec u^\mathrm{\,(i)} = \vec\Psi(\vec q^{\,(0)},t)$ which maps the initial Lagrangian coordinate $\vec q^{\,(0)}$ of any fluid element to its final position $\vec q$ at a later time $t$. (See also \ref{app:A} for the notation.)

Applying this map to the evolution of fluid mass elements leads to the continuity equation
\begin{equation}
  \left(1 + \delta(\vec q,t)\right)\rmd^3q =
  \left(1 + \delta^\mathrm{(i)}\right)\rmd^3q^{(0)} \approx
  \rmd^3q^{(0)}\;,
\label{eq:01-177}
\end{equation}
with $\vec q(t) = \vec q^{\,(0)} + \vec\Psi(\vec q^{\,(0)},t)$, assuming that the initial density field is nearly uniform, $\delta^\mathrm{(i)}\ll1$. Using the Jacobian determinant of this mapping and linearising, one finds the first-order relation
\begin{equation}
  \nabla^{(0)}\cdot\vec\Psi^{(1)}(\vec q^{\,(0)},t) =
  -\delta^{(1)}\left(\vec q^{\,(0)}+\vec\Psi^{(1)}(\vec q^{\,(0)},t),t\right)\;.
\label{eq:01-178}
\end{equation}
Together with the equation of motion for $\vec\Psi$ \cite{2008PhRvD..77f3530M} and assuming an irrotational flow, this equation is solved by \cite{1996MNRAS.282..767T}
\begin{equation}
  \vec\Psi^{(1)}(\vec q^{\,(0)},t) =
  D_+(t)\int_k\,\frac{\rmi\vec k}{k^2}\,\delta^\mathrm{(i)}(\vec k)\,
  \e^{-\rmi\vec k\cdot\vec q^{\,(0)}}\;.
\label{eq:01-179}
\end{equation}
Thus, in this approximation, particles simply move on straight trajectories with the time dependence given by the linear growth factor. The Zel'dovich approximation now lies in extrapolating these trajectories to the present day. In our approach, this entails replacing the Hamiltonian propagator (\ref{eq:01-175}) with a Zel'dovich propagator
\begin{equation}
  g_{qp}^\mathrm{(Z)}(\tau,\tau') = \tau-\tau'\;.
\label{eq:01-180}
\end{equation}
Then, the time-evolution factor in the free two-point density cumulant (\ref{eq:01-130}) derived from linear initial correlations turns into
\begin{equation}
  1 + g_{qp}^\mathrm{(Z)}(\tau_1,0) = 1+\tau_1 = D_+(\tau_1)\;.
\label{eq:01-181}
\end{equation}
Consequently, the free linear power-spectrum contribution scales as $P_\delta \propto D_+^2(a)$, as expected from Eulerian standard perturbation theory (SPT). This time evolution is due to the fact that the equation of motion for $\vec\Psi^{(1)}$ has the same form as that for the linear density contrast in SPT. Since this equation of motion contains the gravitational potential, the Zel'dovich trajectories already include part of the gravitational interaction between particles. This interaction and the actual deviations from inertial motion it causes are hidden in the time dependence of straight Zel'dovich trajectories.

To first and second order in the initial correlations, with the Hamiltonian propagator replaced by the Zel'dovich propagator and suitably approximated damping terms, we thus find the following contributions to the power spectrum in our free theory
\begin{eqnarray}
  P_\delta^{(1)}(k) &=& P_\delta^\mathrm{(i)}\left(k\right)
  \left(1+\tau_1\right)^2 = D_+^2P_\delta^\mathrm{(i)}(k)\;,\nonumber\\
  P_\delta^{(2)}(k) &=& \frac{\tau_1^4}{2\left(
    1+\frac{\sigma_1^2}{3}\tau_1^2k^2
  \right)}
  \int_{k'}
  P_\delta^\mathrm{(i)}\left(k'\right)
  P_\delta^\mathrm{(i)}\left(\vec k-\vec k'\right)
  \left(\frac{\vec k\cdot\vec k'}{k'^2}\right)^2
  \left(
    \frac{\vec k\cdot(\vec k-\vec k')}{(\vec k-\vec k')^2}
  \right)^2\;,
\label{eq:01-182}
\end{eqnarray}
where we have specified for clarity that the power spectra on the right-hand sides are the power spectra $P_\delta^\mathrm{(i)}$ characterising the initial particle distribution.

As one would expect, these expressions are also found in linear LPT by going to quadratic order in the initial correlations of $\vec\Psi^{(1)}$ (cf.\ the $C^{(11)}_{ij}$ terms in equation (35) of \cite{2008PhRvD..77f3530M}). However, one should be aware of the fact that the LPT formalism only includes pure initial momentum correlations due to the assumption of a uniform initial density in (\ref{eq:01-177}). In our approach, quadratic power-spectrum contributions coming from density auto-correlations and density-momentum cross-correlations are also present, as they should be. As mentioned before, we dropped them here since they scale with lower powers of the propagator $g_{qp}$. The assumption of a uniform initial density field is also responsible for the slightly different time dependence of LPT when compared with our approach.

Strictly speaking, the choice of LPT in (\ref{eq:01-179}) is inconsistent with this assumption and the boundary condition $\vec\Psi(\vec q^{\,(0)},0) = 0$ if the linear growth is normalised as $D_+(0) = 1$ at the initial time $t = 0$. However, it is necessary in LPT to achieve the same growth of the linear power spectrum with time as in Eulerian SPT, since LPT lacks the density auto-correlation and the density-momentum cross-correlations which together lead to the correct time evolution factor in (\ref{eq:01-130}) and (\ref{eq:01-181}).

\begin{figure}
  \includegraphics[width=\hsize]{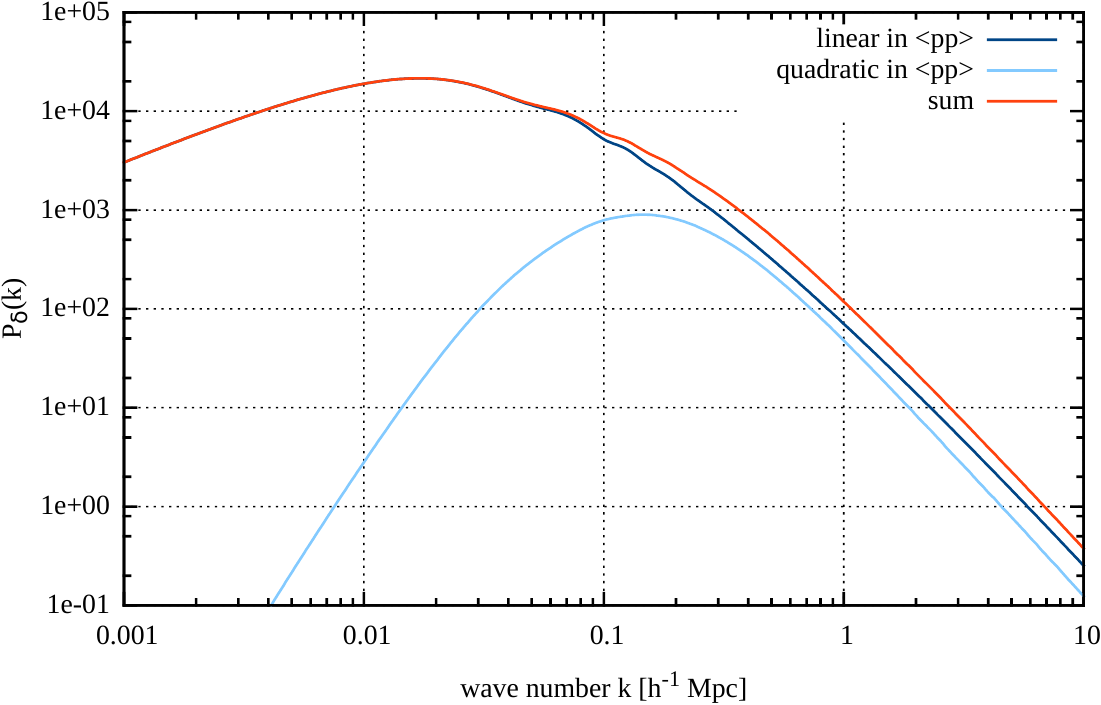}
\caption{The contributions $P_\delta^{(1)}(k)$ and $P_\delta^{(2)}(k)$ to the density power spectrum obtained from the free generating functional with the Zel'dovich propagator are shown for a CDM initial power spectrum, evolved to $\sigma_8 = 0.8$.}
\label{fig:1}
\end{figure}

The two contributions (\ref{eq:01-182}) are shown in Figure~\ref{fig:1} for an initial CDM power spectrum evolved in a standard $\Lambda$CDM universe such that its present normalisation reaches $\sigma_8 = 0.8$.

The integrals remaining in the expressions can quickly and accurately be carried out by Monte-Carlo integration. With the Monte-Carlo integrator contained in the Gnu Science Library, evaluating the integrand in (\ref{eq:01-182}) $10^5$ times, the curves in Figure~\ref{fig:1} require $\sim30$~seconds on a single core of a slightly outdated desktop PC.

\subsection{Bispectrum}
\label{ssc:7.2}

Of the three-point terms $G_{\rho\rho\rho}^{(1)}(123)$, $G_{\rho\rho\rho}^{(2A)}(123)$ and $G_{\rho\rho\rho}^{(2B)}(123)$ in (\ref{eq:01-151}), (\ref{eq:01-153}) and (\ref{eq:01-155}), only $G_{\rho\rho\rho}^{(2B)}(123)$ is a connected contribution to the bispectrum. Interchanging $\vec k_2$ and $\vec k_3$ there, and taking into account that the preceding delta distribution ensures $\vec k_3 = -(\vec k_1+\vec k_2)$, we can write
\begin{equation}
  G_{\rho\rho\rho}^{(2B)}(123) = \bar\rho^3(2\pi)^3
  \delta_\mathrm{D}\left(\vec k_1+\vec k_2+\vec k_3\right)\tau_1^4
  \left\{
    P_\delta\left(\vec k_1\right)P_\delta\left(\vec k_2\right)
    F\left(\vec k_1,\vec k_2\right) + \mbox{cyc.}
  \right\}
\label{eq:01-183}
\end{equation}
with the kernel $F$ defined by
\begin{equation}
  F\left(\vec k_1,\vec k_2\right) := 1+
  \frac{\vec k_1\cdot\vec k_2}{k_1k_2}
  \left(\frac{k_1}{k_2}+\frac{k_2}{k_1}\right)+
  \frac{(\vec k_1\cdot\vec k_2)^2}{k_1^2k_2^2}\;.
\label{eq:01-184}
\end{equation}
This structurally reproduces the $2F_2$-kernel appearing in Eulerian perturbation theory of the density contrast,
\begin{equation}
  2F_2\left(\vec k_1,\vec k_2\right) =
  \frac{10}{7}+\frac{\vec k_1\cdot\vec k_2}{k_1k_2}
  \left(\frac{k_1}{k_2}+\frac{k_2}{k_1}\right)+
  \frac{4}{7}\frac{(\vec k_1\cdot\vec k_2)^2}{k_1^2k_2^2}\;;
\label{eq:01-185}
\end{equation}
(cf.\ Eq.~(43) in \cite{2002PhR...367....1B}). Remaining differences are due to the different levels of self-gravity included here and in Eulerian perturbation theory and will be detailed in a future study.

\subsection{First-order results for non-linear evolution}

In first-order perturbation theory, we have the free contributions $G_{\rho\rho}^{(1)}$ and $G_{\rho\rho}^{(2)}$ from (\ref{eq:01-130}) and (\ref{eq:01-133}) to the power spectrum together, the contribution $\delta^{(1)}G_{\rho\rho}^{(1)}$ from linear momentum correlations given in (\ref{eq:01-167}), and the contribution $\delta^{(2)}G_{\rho\rho}^{(2)}$ from quadratic momentum correlations shown in (\ref{eq:01-174}) with the individual terms listed in (\ref{eq:01-172}).

We have shown in \cite{2015PhRvD..91h3524B} that the Zel'dovich propagator $g_{qp}^\mathrm{(Z)}$ can be improved and replaced by
\begin{equation}
  \tilde g_{qp}(\tau,\tau') :=
  \int_{\tau'}^\tau\rmd\bar\tau\exp\left(h(\bar\tau)-h(\tau')\right)\;,
\label{eq:01-186}
\end{equation}
where the function $h(\tau)$ is given by
\begin{equation}
  h(\tau) := g^{-1}(\tau)-1
\label{eq:01-187}
\end{equation}
in terms of the function $g(\tau)$ from (\ref{eq:01-176}). Since $g(\tau)\to1$ for $\tau\to0$, we have $h(\tau)\to0$ initially.

The potential acting on particles following the improved Zel'dovich trajectories was shown in \cite{2015PhRvD..91h3524B} to satisfy the Poisson equation
\begin{equation}
  \nabla_q^2v = \frac{3}{2}\frac{a}{g^2(a)}\Omega_\mathrm{mi}\delta\;.
\label{eq:01-188}
\end{equation}
As detailed in \cite{2015PhRvD..91h3524B}, it is an important aspect of the improved Zel'dovich approximation that it clarifies what fraction of gravity is captured already by the free trajectories and what gravitational potential acts on these trajectories in addition. Since the matter-density parameter $\Omega_\mathrm{mi}$ is also to be evaluated at the arbitrarily early initial time here, we can set $\Omega_\mathrm{mi} = 1$ for any Friedman cosmology. Following (\ref{eq:01-116}) and (\ref{eq:01-119}), we can then write the Fourier transform of the one-particle potential as
\begin{equation}
  v(1) = -\frac{g_v(\tau_1)}{\bar\rho k_1^2}\;,\quad
  g_v(\tau) = \frac{3}{2}\frac{a}{g^2(a)}\;.
\label{eq:01-189}
\end{equation}

With the potential (\ref{eq:01-189}), the first-order perturbation contributions to the non-linear power spectrum are
\begin{eqnarray}
\label{eq:01-190}
  \fl
  \delta^{(1)}P_\delta^{(1)}(k_1) &=
  2\left(1+\tilde g_{qp}(\tau)\right)P_\delta^\mathrm{(i)}(k_1)
  \int_0^\tau\rmd\tau'
  g_v(\tau')\tilde g_{qp}(\tau,\tau')\left(1+\tilde g_{qp}(\tau')\right)\;,\\
  \fl
  \delta^{(1)}P_\delta^{(2)}(k_1) &= 2\int_0^\tau\rmd\tau'
  g_v(\tau')\tilde g_{qp}(\tau,\tau')
  \int_{k_1'}\frac{\vec k_1\cdot\vec k_1'}{k_1'^2}\left(
    Z_{1223}^{(2A)}+Z_{2331}^{(2B)}+Z_{3112}^{(2C)}+
    \frac{Z_{2323}^{(2D)}}{2}
  \right)\;,\nonumber
\end{eqnarray}
with the terms $Z_{jkkl}^{(2A,B,C)}$ given in (\ref{eq:01-168})-(\ref{eq:01-170}) and $Z_{jkjk}^{(2D)}$ from (\ref{eq:01-171}).

Clearly, the contributions $P_\delta^{(1)}(k_1)$ from (\ref{eq:01-182}, now to be taken with the improved Zel'dovich propagator $\tilde g_{qp}$) and $\delta^{(1)}P_\delta^{(1)}(k_1)$ from (\ref{eq:01-190}) are both proportional to the initial power spectrum, at least for large scales or small wave numbers $k_1$. Together, they reproduce the linear growth of the power spectrum. Since this is simply given by $P_\delta(k_1) = D_+^2P_\delta^\mathrm{(i)}$, we neglect these terms here and focus on the non-linear term $\delta^{(1)}P_\delta^{(2)}(k_1)$ from (\ref{eq:01-190}) with its four contributions.

\begin{figure}[ht]
  \includegraphics[width=\hsize]{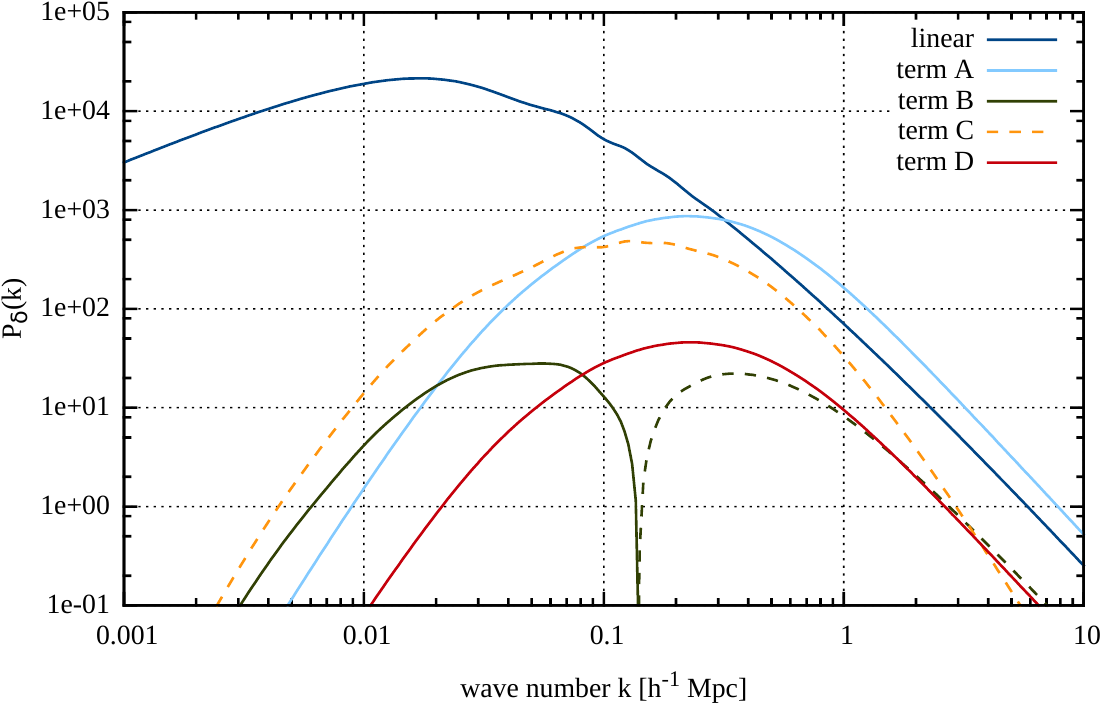}
\caption{Terms A, B, C and D as defined in (\ref{eq:01-168})-(\ref{eq:01-170}) and (\ref{eq:01-171}) contributing at first-order perturbation theory to the non-linear evolution of the density power spectrum. Dashed curves indicate negative terms. For reference, the linearly evolved density power spectrum is also shown, evolved to $\sigma_8 = 0.8$.}
\label{fig:2}
\end{figure}

Figure~\ref{fig:2} shows these four terms A, B, C and D together with the linearly evolved density power spectrum. Dashed curves indicate negative contributions. Clearly, term A is positive throughout and expresses how structures grow on small scales by gravitational collapse. Term C is negative. At large scales, its amplitude is larger than that of term A, while it falls below at small scales. This reflects two important aspects of cosmological structure formation. Gravitational contraction removes power from large scales and transports it to smaller scales. However, the reduction of power on small scales by term C is exaggerated here because the improved Zel'dovich propagators still overshoot and are only partly compensated by the first-order interaction. The power on small scales is thus suppressed too strongly. The terms B and D are much lower in amplitude, except on the smallest scales. There, however, the negative contribution by term B almost exactly cancels the positive contribution by term D. While term D contributes to structure growth on small scales, term B adds power on intermediate scales, but removes power on small scales, albeit at a low level.

\begin{figure}[]
  \includegraphics[width=\hsize]{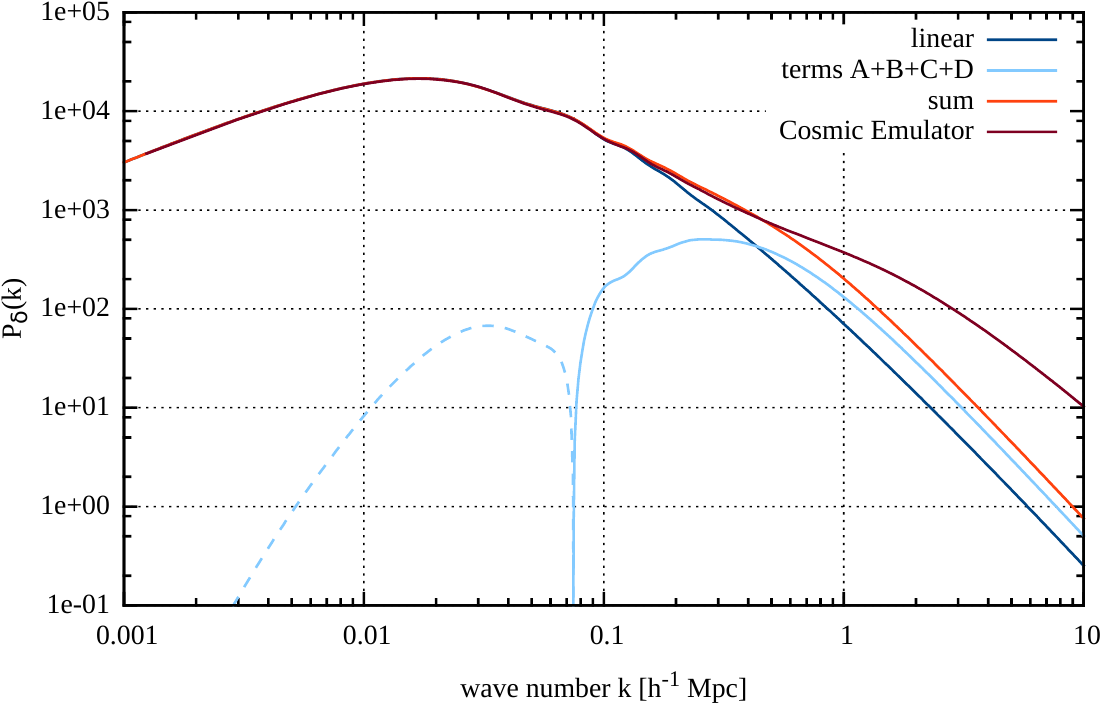}
\caption{The sum of terms A-D is shown together with the linearly evolved power spectrum and added to it. For reference, the cosmic-density power spectrum non-linearly evolved to the present in numerical simulations and reproduced by the Coyote cosmic emulator \cite{2014ApJ...780..111H} is overplotted. The spectra are are evolved to $\sigma_8 = 0.8$ for linear evolution.}
\label{fig:3}
\end{figure}

The combined effect of terms A-D is shown in Figure~(\ref{fig:3}), together with the sum of terms A-D and the linearly evolved power spectrum. For comparison, the approximation by the Coyote cosmic emulator \cite{2014ApJ...780..111H} of the non-linearly evolved density power spectrum obtained in fully numerical simulations is also shown. Clearly, at first order perturbation theory, our analytic result falls below the numerical result. The reduction of power on large scales and the increase on small scales is clearly shown. Yet, the overshooting due to the improved Zel'dovich propagator is not adequately compensated by the first-order gravitational interaction, causing a loss of power on small scales, where structures are wiped out.

\begin{figure}[ht]
  \includegraphics[width=\hsize]{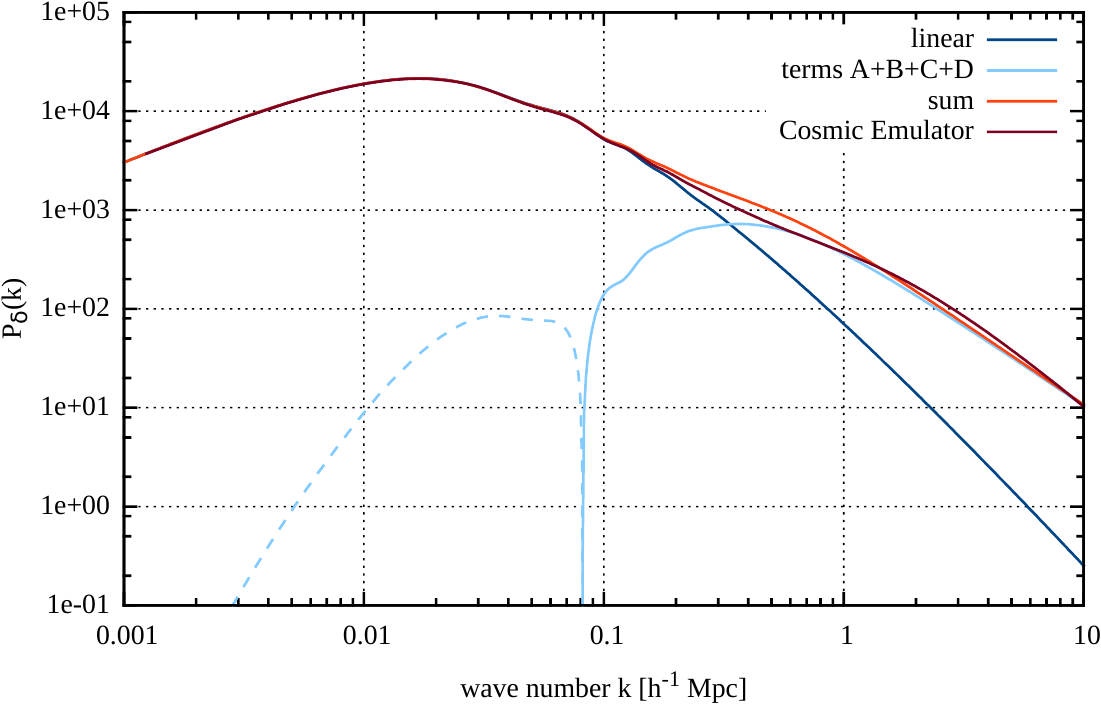}
\caption{The combined contributions A-D to the first-order non-linear density power spectrum are shown, calculated with the modified interaction potential (\ref{eq:01-194}). Also shown is the sum of the linear power spectrum and the terms A-D, and the approximation by the Coyote cosmic emulator \cite{2014ApJ...780..111H} of the numerically simulated density power spectrum. Both the shape and the amplitude of the power spectrum on non-linear scales are well reproduced. The spectra are are evolved to $\sigma_8 = 0.8$ for linear evolution.}
\label{fig:4}
\end{figure}

Clearly, this should be improved by going to second-order perturbation theory, as we shall do in a subsequent paper. There is however, a simple remedy even at first order, following the idea of the adhesion approximation \cite{1989RvMP...61..185S, 1989MNRAS.236..385G, 2012PhyU...55..223G}. This approximation was introduced to compensate the effect of free streaming in the Zel'dovich approximation once particle trajectories have crossed. This is achieved by adding a viscosity term to the otherwise inertial motion in the Zel'dovich approximation,
\begin{equation}
  \frac{\rmd\vec v}{\rmd\tau} = 0\quad\to\quad
  \frac{\rmd\vec v}{\rmd\tau} = \nu\nabla^2\vec v\;,
\label{eq:01-191}
\end{equation}
where the viscosity needs to be adequately adapted. Since the velocity is initially the gradient of a velocity potential $\psi$, $\vec v = \nabla\psi$, whose Laplacian is the negative density contrast, $\nabla^2\psi = -\delta$, the right-hand side of (\ref{eq:01-191}) corresponds to a force proportional to the negative gradient of the density contrast $\delta$. This is quite intuitive: particles will be kept near steep density gradients. This force can be introduced by adding a term proportional to the density contrast to the potential. The potential (\ref{eq:01-119}) would then change to
\begin{equation}
  v(1) = -\frac{g_v(\tau_1)}{\bar\rho}\left(\frac{1}{k_1^2}+\nu\right)\;.
\label{eq:01-192}
\end{equation}
However, the adhesion approximation is known to yield razor-sharp dark-matter filaments which are considerably narrower than those found in numerical simulations. They can be softened replacing (\ref{eq:01-192}) by
\begin{equation}
  v(1) = -\frac{g_v(\tau_1)}{\bar\rho}
  \left(\frac{1}{k_1^2}+\frac{\bar\nu}{k_1}\right)\;,
\label{eq:01-193}
\end{equation}
where $\bar\nu$ is an amplitude with the dimension of a length scale. As a suitable length scale, we choose twice the velocity dispersion $\sigma_v$, propagated to the time $\tau$ by the improved Zel'dovich propagator $\tilde g_{qp}(\tau, 0)$,
\begin{equation}
  \bar\nu = \tilde g_{qp}(\tau, 0)\,\sigma_v\;;
\label{eq:01-194}
\end{equation}
note that $\tilde g_{qp}(\tau, 0)\,\sigma_v\approx2$ at late times. Inserting this modified interaction potential into our formalism, we obtain the non-linear power spectrum shown in Figure~\ref{fig:4}. We emphasise that we introduce the adhesion approximation and set the viscosity parameter to overcome the limitations of the first-order perturbation theory. We expect that, once we proceed to higher perturbative orders, the theory will be parameter-free.

As Fig.~\ref{fig:4} shows, the amplitude and the shape of our non-linear density power spectrum at first-order perturbation theory now agrees very well with the approximation by the Coyote cosmic emulator \cite{2014ApJ...780..111H} of the fully numerical density power spectrum.

\begin{figure}[ht]
  \includegraphics[width=\hsize]{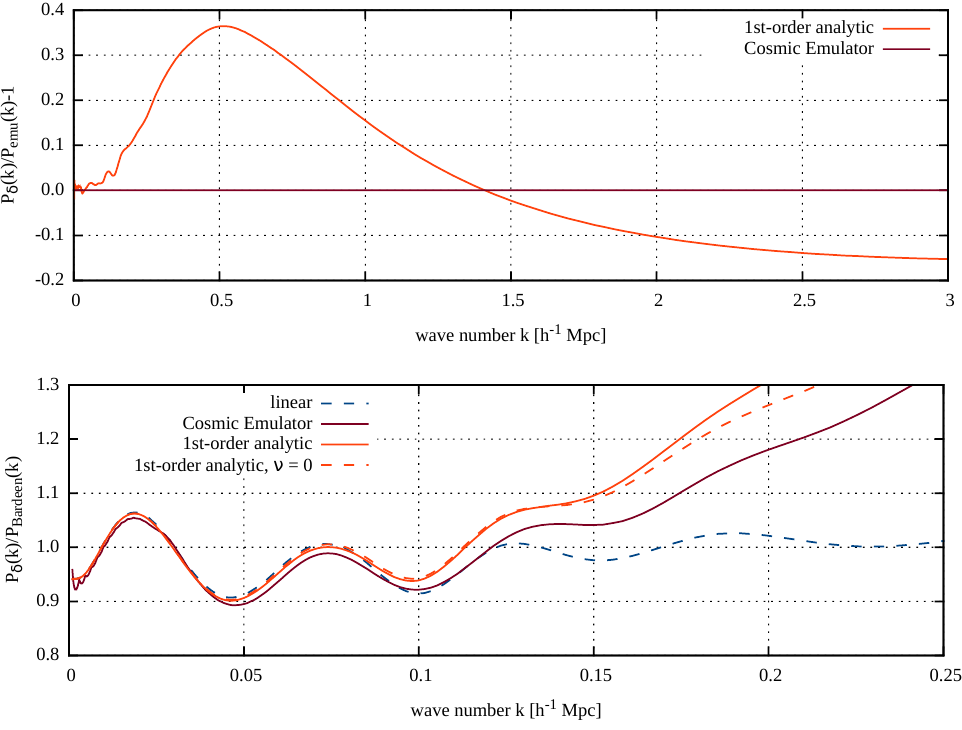}
\caption{\emph{Upper panel}: Linear plot of the region in $k$ space where the relative difference between our first-order analytic power spectrum from the result of the Cosmic Emulator is largest. The relative difference reaches $\approx35\,\%$ near $k\approx0.5\,h\,\mathrm{Mpc}^{-1}$ and drops to $\approx-15\,\%$ near $k\approx3\,h\,\mathrm{Mpc}^{-1}$. \emph{Lower panel}: Linear plot of the region in $k$ space where the baryonic acoustic oscillations (BAOs) are most pronounced. To reduce the dynamic range, we normalize the linearly evolved power spectrum, the power spectrum evolved with the Cosmic Emulator and our first-order analytic solutions with and without viscosity to the linear spectrum according to \cite{1986ApJ...304...15B}, which has no BAOs. The spectra are are evolved to $\sigma_8 = 0.8$ for linear evolution.}
\label{fig:5}
\end{figure}

To emphasise the remaining differences between our first-order result and the density power spectrum found in numerical simulations, we show the difference between our result and the power spectrum according to the Coyote cosmic emulator with linear scaling in the upper panel of Fig.~\ref{fig:5}. The lower panel shows how the baryonic acoustic oscillations are reproduced by our approach.

Of course, these are not final answers yet, but results which we believe to be very encouraging. Even at first order in the perturbation theory, with a modification of the interaction potential modelled after the adhesion approximation, the non-linear power spectrum comes out very close to fully numerical result. This suggests that second-order perturbation theory may already yield fully satisfactory results with the Newtonian interaction potential (\ref{eq:01-189}). This will be worked out in a follow-up study.

\section{Conclusions}
\label{sc:8}

Mazenko and Das \& Mazenko have recently described how the evolution of classical point-particle ensembles can be described as a non-equilibrium statistical field theory \cite{2010PhRvE..81f1102M, 2012JSP...149..643D}. The theory uses the phase-space coordinates of the $N$ particles in the ensemble as elementary microscopic fields. The central object of the theory is a generating functional. This is a path integral over phase-space trajectories, weighed by a probability distribution for the initial particle positions, in which each particle is represented by a phase factor containing the free particle trajectory.

For expressing the microscopic properties of the $N$-particle ensemble in a compact way, we have introduced a notation bundling these properties in tensor-valued structures which appear straightforward to calculate with (see also \cite{1986ApJ...304...15B}).

Collective fields, such as the density and the response field, are described by operators extracting correlators of the collective fields from the generating functional by functional derivatives. Likewise, interactions between particles are described by an interaction operator containing the collective density and response fields. This theory closely resembles statistical quantum field theory. Since the form for the equation of motion of the microscopic degrees of freedom is assumed to be very general, the theory should be widely applicable to ensembles of classical point particles.

Using the theory of cosmic structure formation as a motivation, we have derived in Sect.~\ref{sc:4} a probability distribution for the initial phase-space positions of the particles which accounts for auto-correlations of the spatial positions and the momenta, as well as for cross-correlations between spatial positions and momenta.

We argue that, under very general conditions, continuity requires the momenta to be correlated if the positions are. If the particles are supposed to sample an initial Gaussian density field, a single initial power spectrum, e.g.\ for the initial density field, suffices to specify the initial probability distribution. We derive its exact form, which contains a correlation operator containing the complete correlation hierarchy of the ensemble. We then approximate this correlation operator to low order in the correlations and give explicit expressions for the generating functional containing momentum correlations to first and second order.

The main results of this derivation in Sect.~\ref{sc:4} are the exact equation (\ref{eq:01-65}) for the initial probability distribution in phase space, the contributions (\ref{eq:01-102}) to the generating functional with linear momentum correlations, and the terms (\ref{eq:01-102}), (\ref{eq:01-107}) and (\ref{eq:01-108}) for the generating functional with quadratic momentum correlations. From these results, low-order density- and response-field correlators for correlated ensembles of classical particles can be readily determined, as shown in Sect.~\ref{sc:6}.

Based on this general formalism of the theory and on a free generating functional for initially correlated, canonical particle ensembles, we have derived two- and three-point correlators of the cosmic density field without interaction, and the two-point density correlator of the cosmic density field with interaction included in first-order perturbation theory. Our results of Sect.~\ref{sc:7} can be summarised as follows:

\begin{itemize}
  \item If momentum correlations are taken into account to linear order, and if the free particle propagator is taken from the Zel'dovich approximation, the density power spectrum (\ref{eq:01-130}) reproduces the linear growth well known from standard perturbation theory.
  \item Evolving quadratic momentum correlations with the free Zel'dovich propagator leads to a first contribution to the nonlinear evolution of the power spectrum, for which the simple, closed expression (\ref{eq:01-133}) can be given. This contribution is a convolution of the initial power spectrum with itself, multiplied by a mode-coupling kernel.
  \item Deriving the bispectrum, we obtain the connected term (\ref{eq:01-183}), containing the kernel (\ref{eq:01-184}). In this form, it resembles the bispectrum result from Eulerian perturbation theory, but with a small difference in two coefficients. The reasons for this difference are subtle and will be explained in future work.
  \item Proceeding to first-order perturbation theory, using the improved Zel'dovich propagator and the appropriate interaction potential derived in \cite{2015PhRvD..91h3524B}, we showed that the shape of the first-order non-linear terms reproduce the shape of the non-linear density power spectrum known from numerical simulations rather well, while the amplitude at large wave numbers turns out to be substantially too low. This reflects the fact that, in the improved Zel'dovich approximation, the re-expansion of cosmic structures is still not fully suppressed by the first-order interaction.
  \item While this calls for higher-order perturbation theory, an effective remedy is provided even at first order by an adapted version of the adhesion approximation \cite{1989RvMP...61..185S, 1989MNRAS.236..385G, 2012PhyU...55..223G}. If we modify the interaction potential accordingly, the shape as well as the amplitude of the non-linear corrections to the power spectrum reproduce the numerical results very well.
\end{itemize}

Our calculation extends to redshift zero and to arbitrary wave numbers. Apart from the viscosity introduced to strengthen gravity in our first-order calculation, our theory has no free parameters once the power spectrum of the initial phase-space particle distribution is fixed and normalised. We expect that, once we can proceed to second-order perturbations, the theory will have no free parameters. The form of the non-linear terms and the inevitable damping factor suggest that the expected asymptotic behaviour of the non-linear power spectrum for large wave numbers will be retained in higher-order calculations.

The main difference to conventional, Eulerian or Lagrangian perturbation theory of cosmic-structure evolution is that we do not require, solve or perturb a dynamical equation for the cosmic density. Rather, we study the statistical evolution of a particle ensemble in phase space, weakly perturbing their trajectories, and read out any collective information, such as the density, from the evolved phase-space distribution when needed. Since even small perturbations of trajectories can lead to large increases in density, our approach is able to extend into the regime of highly non-linear density perturbations even at low perturbative orders. It also appears crucial to keep the complete phase-space information of the particles because this allows us to use the Hamiltonian equations of motion with their simple structure and their equally simple Green's function.

While our main motivation for this work has been the extension of this theory to cosmic structure formation from dark-matter particles, the results given here may be useful for a wider class of problems involving the non-equilibrium statistics of correlated ensembles of classical particles.

\ack

We wish to thank Luca Amendola, J\"urgen Berges, Manfred Salmhofer, Bj\"orn Sch\"afer and Christof Wetterich for inspiring and helpful discussions. This work was supported in part by the Transregional Collaborative Research Centre TR~33, ``The Dark Universe'', of the German Science Foundation (DFG) and by the Munich Institute for Astro- and Particle Physics (MIAPP) of the DFG cluster of excellence ``Origin and Structure of the Universe''.

\appendix

\section{Probability distribution for initial particle positions and momenta}
\label{app:A}

\subsection{Velocity potential}

We assume that the particles are initially located at spatial positions $\vec q^\mathrm{\,(0)}$, slightly displaced by an initial displacement field $\vec u^\mathrm{\,(i)}$. The particle trajectories near the initial time $t = 0$ are
\begin{equation}
  \vec q(t) = \vec q^\mathrm{\,(0)}+b(t)\vec u^\mathrm{\,(i)}
\label{eq:A01-1}
\end{equation}
with a yet unknown monotonic function of time $b(t)$ which does not need to be further specified for now. Without loss of generality, we set $b(t=0) = 1$.

If the initial particle velocities sample an irrotational velocity field, a velocity potential $\psi$ exists such that
\begin{equation}
  \vec u^\mathrm{\,(i)} = \partial_q^\mathrm{(0)}\psi\;.
\label{eq:A01-2}
\end{equation}
Then, continuity implies that the density evolves as
\begin{equation}
  \rho(t) = \bar\rho\left\vert
    \det\left(\frac{\partial\vec q(t)}{\partial\vec q^\mathrm{\,(0)}}\right)
  \right\vert^{-1} =
  \bar\rho\left\vert\det\left(
    \delta_{ij}+b(t)
    \frac{\partial^2\psi}
    {\partial q_i^\mathrm{(0)}\partial q_j^\mathrm{(0)}}
  \right)\right\vert^{-1}\;.
\label{eq:A01-3}
\end{equation}
If the elements of the Hessian of $\psi$ are small, the determinant can be expanded,
\begin{equation}
  \rho(t) \approx \bar\rho\left(
    1+b(t)\mathrm{tr}
    \frac{\partial^2\psi}
    {\partial q_i^\mathrm{(0)}\partial q_j^\mathrm{(0)}}
  \right)^{-1} \approx \bar\rho\left(
    1-b(t)\left(\nabla^\mathrm{(0)}\right)^2\psi
  \right)\;.
\label{eq:A01-4}
\end{equation}
implying that the initial density contrast needs to satisfy the Poisson equation
\begin{equation}
  \delta^\mathrm{(i)} = -\left(\nabla^\mathrm{(0)}\right)^2\psi\;.
\label{eq:A01-5}
\end{equation}
The power spectra $P_\psi(k)$ for the velocity potential and $P_\delta(k)$ for the density contrast must thus be related by
\begin{equation}
  P_\psi(k) = k^{-4}P_\delta(k)\;.
\label{eq:A01-6}
\end{equation}
At the same time, the initial particle velocity is
\begin{equation}
  \dot{\vec q}^\mathrm{\,(i)} = \dot b(t=0)\vec u^\mathrm{\,(i)} =
  \dot b(t=0)\nabla^\mathrm{(0)}\psi\;.
\label{eq:A01-7}
\end{equation}
Finally choosing the time coordinate such that $\dot b(t=0) = 1$ and setting the particle mass to unity, we can identify the gradient of the velocity potential with the initial momentum,
\begin{equation}
  \vec p^\mathrm{\,(i)} = \nabla^\mathrm{(0)}\psi\;.
\label{eq:A01-8}
\end{equation}
The initial density contrast and the initial momentum are thus related by the velocity potential. For the remainder of \ref{app:A}, we shall drop the superscript (i) for the initial positions and momenta, understanding that all positions and momenta are to be taken at the initial time for now.

\subsection{Probability for particle positions in phase space}

We aim at the probability for finding a point particle $j$ at position $\vec q_j$ with momentum $\vec p_j$. The point particles need to sample the density $\rho$, hence the probability for finding a particle at position $\vec q_j$, given the density $\rho_j = \rho(\vec q_j)$, must be proportional to $\rho_j$,
\begin{equation}
  P(\vec q_j\,\vert\,\rho_j) = N^{-1}\rho_j\;,
\label{eq:A01-9}
\end{equation}
with the normalisation $N^{-1}$ set by the requirement $\int\rho_j\rmd V = N$.

Expressing the initial density $\rho$ by the density contrast $\delta$ and introducing the velocity potential $\psi$, we can substitute $\rho = \bar\rho\left(1-\nabla^2\psi\right)$ for the density. Note that $\rho$ must be chosen to be a number density here since we intend to draw particles from it.

In a similar manner, we need to account for the conditional probability for a particle at position $\vec q_j$ to have momentum $\vec p_j$. By (\ref{eq:A01-8}), we have
\begin{equation}
  P\left(\vec p_j\,\vert\,\nabla\psi_j\right) =
  \delta_\mathrm{D}\left(
    \vec p_j-\nabla\psi_j
  \right)\;,
\label{eq:A01-10}
\end{equation}
understanding that $\nabla\psi_j = \nabla\psi(\vec q_j)$. For ease of notation, we now abbreviate the negative Laplacian and the gradient of the potential $\psi$ at position $\vec q_j$ by
\begin{equation}
  \delta_j := -\nabla^2\psi_j\;,\quad
  \vec y_j := \nabla\psi_j\;.
\label{eq:A01-11}
\end{equation}
The probability for finding a particle at position $\vec q_j$ with momentum $\vec p_j$ can thus be related by
\begin{eqnarray}
  P(\vec q_j, \vec p_j) &= \int\rmd\delta_j\int\rmd^3y_j\,
  P(\vec q_j\vert\delta_j)P(\vec p_j\vert\vec y_j)P(\delta_j,\vec p_j)
  \nonumber\\ &=
  \frac{\bar\rho}{N}\int\rmd\delta_j\,(1+\delta_j)\int\rmd^3y_j\,
  \delta_\mathrm{D}\left(\vec p_j-\vec y_j\right)
  P(\delta_j, \vec y_j) \nonumber\\ &=
  \frac{\bar\rho}{N}\int\rmd\delta_j\,(1+\delta_j)
  P\left(\delta_j, \vec p_j\right)
\label{eq:A01-12}
\end{eqnarray}
to the probability $P(\delta_j,\vec p_j)$ for the (negative) Laplacian and the gradient of the velocity potential.

Since the velocity potential is supposed to be a Gaussian random field, its derivatives will also be Gaussian random fields. At each point $\vec q_j$, we will thus have four Gaussian random variables $(\delta_j, \vec p_j)=:(\delta, \vec p\,)_j$, viz.\ the negative Laplacian $\delta_j$ and the gradient $\vec p_j$ of the velocity potential.

We are searching for the probability distribution for a set $\{\vec q_j, \vec p_j\}$ of $N$ particle positions and momenta. It will be given by
\begin{equation}
  P\left(\left\{\vec q_j, \vec p_j\,\right\}\right) = A
  \int\rmd^N\delta\prod_{j=1}^N(1+\delta_j)
    P\left(\left\{\delta_j, \vec p_j\,\right\}\right)\;.
\label{eq:A01-13}
\end{equation}
The normalisation constant $A$ is related to the normalisation in (\ref{eq:A01-9}) by
\begin{equation}
  A := \left(N^{-1}\bar\rho\right)^N = V^{-N}\;.
\label{eq:A01-14}
\end{equation}

\subsection{Covariance matrix}

Given the data tensor
\begin{equation}
  \bi d := \cvector{\delta\\ \vec p}_j\otimes\vec e_j
\label{eq:A01-15}
\end{equation}
defined in the main part of the paper, the covariance matrix of these data can be decomposed as
\begin{eqnarray}
  \bar C &= \left\langle\bi d\otimes\bi d\right\rangle =
  \matrix{cc}
   {\langle\delta_j\delta_k\rangle &
    \left\langle\delta_j\vec p_k\,\right\rangle^\top \\
    \left\langle\delta_k\vec p_j\,\right\rangle &
    \left\langle\vec p_j\otimes\vec p_k\,\right\rangle}
  \otimes E_{jk} \nonumber\\ &=
  C_D\otimes I_N+\sum_{j\ne k}C_{jk}\otimes E_{jk}
\label{eq:A01-16}
\end{eqnarray}
with
\begin{equation}
  E_{jk} := \vec e_j\otimes\vec e_k\;.
\label{eq:A01-17}
\end{equation}
The $4\times4$-dimensional matrix
\begin{equation}
  C_D = \matrix{cc}
   {\langle\delta\delta\rangle &
    \langle\delta\vec p\,\rangle^\top \\
    \langle\delta\vec p\,\rangle &
    \left\langle\vec p\otimes\vec p\,\right\rangle\\}
\label{eq:A01-18}
\end{equation}
contains the one-point variances, which are of course equal for all particles in a statistically homogenous distribution.
The $4\times4$-dimensional matrices
\begin{equation}
  C_{jk} = \matrix{cc}
   {\langle\delta_j\delta_k\rangle &
    \langle\delta_j\vec p_k\,\rangle^\top \\
    \langle\delta_k\vec p_j\,\rangle &
    \left\langle\vec p_j\otimes\vec p_k\,\right\rangle\\}
\label{eq:A01-19}
\end{equation}
contain the two-point variances between different points $\vec q_j$ and $\vec q_k$.

We begin working out the one- and two-point variances defining
\begin{equation}
  \sigma_n^2 := \int_kk^{2n}P_\psi(k) = \int_kk^{2(n-2)}P_\delta(k)
\label{eq:A01-20}
\end{equation}
in terms of the power spectrum $P_\psi(k)$ for the velocity potential $\psi$,
or the power spectrum $P_\delta(k)$ of the density contrast. By definition of
the density-fluctuation power spectrum, we must have
\begin{eqnarray}
  \left\langle\delta_j\delta_k\right\rangle &=
  \int_k\,P_\delta(k)\,
  \e^{-\rmi\vec k\cdot(\vec q_j-\vec q_k)} = \xi_{\delta\delta}(r_{jk})\;,
\label{eq:A01-21}
\end{eqnarray}
with the density correlation function $\xi_{\delta\delta}$ evaluated at $r_{jk} = |\vec q_j-\vec q_k|$. Evidently, for $j=k$, $\xi_{\delta\delta}(r_{jk}) = \xi_{\delta\delta}(0) = \sigma_2^2$.

Similarly, we obtain
\begin{equation}
  \left\langle\delta_j\vec p_k\,\right\rangle =
  \rmi\int_k\,k^2\vec k\,P_\psi(k)\,
  \e^{-\rmi\vec k\cdot\vec r_{jk}}\;.
\label{eq:A01-22}
\end{equation}
Finally, we have
\begin{equation}
  \left\langle\vec p_j\otimes\vec p_k\right\rangle =
  \int_k\,\vec k\otimes\vec k\,P_\psi(k)\,
  \e^{-\rmi\vec k\cdot\vec r_{jk}}\;.
\label{eq:A01-23}
\end{equation}
Clearly, if $j=k$,
\begin{equation}
  \left\langle\delta_j\vec p_k\,\right\rangle = 0 \quad\mbox{and}\quad
  \left\langle\vec p_j\otimes\vec p_k\,\right\rangle =
  \frac{\sigma_1^2}{3}\mathcal{I}_3\;,
\label{eq:A01-24}
\end{equation}
which allows us to write the one-point variances as
\begin{equation}
  C_D = \matrix{cc}{\sigma_2^2 & 0 \\ 0 & \frac{\sigma_1^2}{3}\mathcal{I}_3}\;.
\label{eq:A01-25}
\end{equation}
Regarding the two-point variances, we decompose the matrices $C_{jk}$ as
\begin{equation}
  C_{jk} = \matrix{cc}
   {C_{\delta_j\delta_k} & C_{\delta_jp_k}^\top \\
    C_{\delta_jp_k} & C_{p_jp_k}} =
  \matrix{cc}
   {\left\langle\delta_j\delta_k\right\rangle &
    \left\langle\delta_j\vec p_k\right\rangle^\top \\
    \left\langle\delta_j\vec p_k\right\rangle &
    \left\langle\vec p_j\otimes\vec p_k\right\rangle}\;.
\label{eq:A01-26}
\end{equation}

\subsection{Initial phase-space probability distribution}

Instead of evaluating $P(\bi d)$ directly, we rather evaluate its characteristic function
\begin{equation}
  \Phi_{\bi d}\left(\bi t\right) = \exp\left(
    -\frac{1}{2}\bi t^\top\bar C\bi t
  \right)\;,
\label{eq:A01-28}
\end{equation}
inspired by \cite{1995MNRAS.272..447R}. The tensor $\bi t$ is Fourier-conjugate to the data tensor $\bi d$. We write $\bi t$ in the form
\begin{equation}
  \bi t = \cvector{t_\delta\\\vec t_p}_j\otimes\vec e_j\;,
\label{eq:A01-29}
\end{equation}
where $(t_\delta, \vec t_p)_j$ is Fourier-conjugate to $(\delta, \vec p\,)_j$.

It will turn out to be convenient for our current purposes to define
\begin{equation}
  \bi t_\delta := t_{\delta_j}\otimes\vec e_j\quad\mbox{and}\quad
  \bi t_p := \vec t_{p_j}\otimes\vec e_j
\label{eq:A01-30}
\end{equation}
and to decompose the quadratic form in (\ref{eq:A01-28}) as
\begin{equation}
  \bi t^\top\bar C\bi t =
  \bi t_\delta^\top\bar C_{\delta\delta}\bi t_\delta +
  2\bi t_\delta C_{\delta p}\bi t_p +
  \bi t_p^\top\bar C_{pp}\bi t_p
\label{eq:A01-31}
\end{equation}
with
\begin{eqnarray}
  \bar C_{\delta\delta} &:= \sigma_2^2\otimes\mathcal{I}_N+
  C_{\delta_j\delta_k}\otimes E_{jk}\;,\quad
  C_{\delta p} := C_{\delta_jp_k}\otimes E_{jk}\;, \nonumber\\
  \bar C_{pp} &:= \frac{\sigma_1^2}{3}\mathcal{I}_3\otimes\mathcal{I}_N+
  C_{p_jp_k}\otimes E_{jk}\;.
\label{eq:A01-32}
\end{eqnarray}

The probability distribution for the data tensor is then given by the inverse Fourier transform
\begin{eqnarray}
  P\left(\bi d\right) &=
  \int\frac{\rmd\bi t_p}{(2\pi)^{3N}}\exp\left(
    -\frac{1}{2}\bi t_p^\top\bar C_{pp}\bi t_p+
    \rmi\left\langle\bi t_p,\bi p\right\rangle
  \right)\nonumber\\ &\cdot
  \int\frac{\rmd\bi t_\delta}{(2\pi)^N}\exp\left(
    -\frac{1}{2}\bi t_\delta^\top\bar C_{\delta\delta}\bi t_\delta-
    \bi t_\delta^\top C_{\delta p}\bi t_p+
    \rmi\left\langle\bi t_\delta,\bdelta\right\rangle
  \right)\;,
\label{eq:A01-33}
\end{eqnarray}
where we have implicitly defined
\begin{equation}
  \bdelta := \delta_j\otimes\vec e_j\;,
\label{eq:A01-34}
\end{equation}
recalled the definition of $\bi p$ in (\ref{eq:01-23}) and used the scalar product $\langle\bi a,\bi b\rangle$ defined in (\ref{eq:01-25}).

\subsubsection{Particle distribution in phase space}

According to (\ref{eq:A01-13}), we need to integrate expression (\ref{eq:A01-33}) over all $\delta_j$ and evaluate
\begin{equation}
  I_1(\bi t_\delta) := \int\rmd\bdelta\,
  \prod_{j=1}^N(1+\delta_j)\exp\left(
    \rmi\left\langle\bi t_\delta,\bdelta\right\rangle
  \right)
\label{eq:A01-35}
\end{equation}
first. We substitute $z_j:=1+\delta_j$, $\rmd z_j=\rmd\delta_j$, and find
\begin{equation}
  I_1(\bi t_\delta) =
  \exp\left(-\rmi\sum_{j=1}^Nt_{\delta_j}\right)
  \prod_{j=1}^N\left(\int\rmd z_jz_j\exp\left(\rmi t_{\delta_j}z_j\right)\right)\;.
\label{eq:A01-36}
\end{equation}
The factors under the product are
\begin{equation}
  \int\rmd z_jz_j\exp\left(\rmi t_{\delta_j}z_j\right) =
  -\rmi\frac{\partial}{\partial t_{\delta_j}}\int\rmd z_j
  \exp\left(\rmi t_{\delta_j}z_j\right) =
  -2\pi\rmi\frac{\partial}{\partial t_{\delta_j}}
  \delta_\mathrm{D}(t_{\delta_j})\;,
\label{eq:A01-37}
\end{equation}
leading us to
\begin{equation}
  I_1(\bi t_\delta) =
  (-2\pi\rmi)^N\exp\left(-\rmi\sum_{j=1}^Nt_{\delta_j}\right)
  \left(\prod_{j=1}^N\frac{\partial}{\partial t_{\delta_j}}\right)\,
  \delta_\mathrm{D}\left(\bi t_\delta\right)\;.
\label{eq:A01-38}
\end{equation}
Next, we need to integrate
\begin{equation}
  I_2(\bi t_p) := \int\frac{\rmd\bi t_\delta}{(2\pi)^N}\exp\left(
    -\frac{1}{2}\bi t_\delta^\top\bar C_{\delta\delta}\bi t_\delta-
    \bi t_\delta^\top C_{\delta p}\bi t_p
  \right)I_1(\bi t_\delta)\;.
\label{eq:A01-39}
\end{equation}
After a partial integration to move the derivatives with respect to $t_{\delta_j}$ away from the delta distribution in $I_1(\bi t_\delta)$, we arrive at
\begin{eqnarray}
  I_2(\bi t_p) &=
  \rmi^N\left(\prod_{j=1}^N\frac{\partial}{\partial t_{\delta_j}}\right)
  \left.\exp\left(
    -\frac{1}{2}\bi t_\delta^\top\bar C_{\delta\delta}\bi t_\delta -
    \bi t_\delta^\top C_{\delta p}\bi t_p-\rmi\sum_{j=1}^Nt_{\delta_j}
  \right)\right\vert_{\bi t_\delta=0} \nonumber\\ &=
  \mathcal{C}\left(\bi t_p\right)\;,
\label{eq:A01-40}
\end{eqnarray}
where $\mathcal{C}(\bi t_p)$ is a correlation operator to be evaluated.

Since the argument of the exponential in (\ref{eq:A01-40}) is quadratic in the $t_{\delta_j}$, the derivative operator in (\ref{eq:A01-40}) can only act up to two times. We have
\begin{eqnarray}
  \left.\frac{\partial}{\partial t_{\delta_j}}\left(
    -\frac{1}{2}\bi t_\delta^\top\bar C_{\delta\delta}\bi t_\delta -
    \bi t_\delta^\top C_{\delta p}\bi t_p-\rmi\sum_{j=1}^Nt_{\delta_j}
  \right)\right\vert_{\bi t_\delta=0} &=
  -\rmi\left(1-\rmi C_{\delta_jp_k}\bi t_{p_k}\right)\;,\nonumber\\
  \left.\frac{\partial^2}{\partial t_{\delta_j}\partial t_{\delta_k}}\left(
    -\frac{1}{2}\bi t_\delta^\top\bar C_{\delta\delta}\bi t_\delta -
    \bi t_\delta^\top C_{\delta p}\bi t_p+\rmi\sum_{j=1}^Nt_{\delta_j}
  \right)\right\vert_{\bi t_\delta=0} &= -\bar C_{\delta_j\delta_k}
\label{eq:A01-41}
\end{eqnarray}
and thus obtain the hierarchy
\begin{eqnarray}
  \mathcal{C}(\bi t_p) &= \prod_{j=1}^N\left(
    1-\rmi C_{\delta_jp_k}\bi t_{p_k}
  \right) +
  \sum_{(j,k)}C_{\delta_j\delta_k}\prod_{\{l\}'}\left(
    1-\rmi C_{\delta_lp_k}\bi t_{p_k}
  \right)\nonumber\\ &+
  \sum_{(j,k)}C_{\delta_j\delta_k}\sum_{(a,b)'}C_{\delta_a\delta_b}
  \prod_{\{l\}''}\left(1-\rmi C_{\delta_lp_k}\bi t_{p_k}\right) + \ldots
\label{eq:A01-42}
\end{eqnarray}
for the correlation operator. In the first line, the sum extends over all pairs $(j,k\ne j)$, and the product includes all indices $l$ except $(j,k)$. In the second line, the sum extends over products of pairs $(j,k\ne j)$ and $(a,b\ne a)$, where $(a,b)'$ excludes the indices $j$ and $k$, and the product includes all indices $l$ except $(j,k,a,b)$. Analogous terms with products over three, four and more pairs have to be added.

Now, we have to insert $I_2(\bi t_p) = \mathcal{C}(\bi t_p)$ into (\ref{eq:A01-33}) and to integrate
\begin{equation}
  I_3(\bi p) := \int\frac{\rmd\bi t_p}{(2\pi)^{3N}}\,\mathcal{C}(\bi t_p)
  \exp\left(-\frac{1}{2}\bi t_p^\top\bar C_{pp}\bi t_p+\rmi\left\langle\bi t_p,\bi p\right\rangle\,\right)\;.
\label{eq:A01-43}
\end{equation}
Here, we can express the remaining $\bi t_p$ factors in the correlation operator by derivatives with respect to $\bi p$, write
\begin{eqnarray}
  \mathcal{C}(\bi p) &= \prod_{j=1}^N\left(
    1-C_{\delta_jp_k}\frac{\partial}{\partial p_k}
  \right) +
  \sum_{(j,k)}C_{\delta_j\delta_k}\prod_{\{l\}'}\left(
    1-C_{\delta_lp_k}\frac{\partial}{\partial p_k}
  \right) \nonumber\\ &+
  \sum_{(j,k)}C_{\delta_j\delta_k}\sum_{(a,b)'}C_{\delta_a\delta_b}
  \prod_{\{l\}''}\left(
    1-C_{\delta_lp_k}\frac{\partial}{\partial p_k}
  \right) + \ldots
\label{eq:A01-44}
\end{eqnarray}
and pull the correlation operator out of the integral to find
\begin{eqnarray}
  I_3(\bi p) &= \mathcal{C}(\bi p)\int\frac{\rmd\bi t_p}{(2\pi)^{3N}}\,
  \exp\left(
    -\frac{1}{2}\bi t_p^\top\bar C_{pp}\bi t_p+
    \rmi\left\langle\bi t_p,\bi p\right\rangle\,
  \right) \nonumber\\ &=
  \frac{1}{\sqrt{(2\pi)^{3N}\det\bar C_{pp}}}\,\mathcal{C}(\bi p)
  \exp\left(-\frac{1}{2}\bi p^\top\bar C_{pp}^{-1}\bi p\,\right)\;.
\label{eq:A01-45}
\end{eqnarray}
Given this result, we finally obtain from (\ref{eq:A01-13}) the probability distribution
\begin{equation}
  P(\bi q, \bi p) = \frac{V^{-N}}{\sqrt{(2\pi)^{3N}\det\bar C_{pp}}}\,
  \mathcal{C}(\bi p\,)
  \exp\left(-\frac{1}{2}\bi p^\top\bar C_{pp}^{-1}\bi p\right)\;,
\label{eq:A01-46}
\end{equation}
with $\bi q$ as defined in (\ref{eq:01-23}).

\bibliographystyle{iopart-num}
\bibliography{../bibliography/main}

\end{document}